\documentclass[longauth]{aaEC}

\usepackage{graphicx}
\usepackage{natbib}
\usepackage{scalerel}
\usepackage[table]{xcolor}
\usepackage{txfonts}
\usepackage[pdfencoding=auto,psdextra]{hyperref}
\hypersetup{
    colorlinks=true,
    linkcolor=blue,
    filecolor=magenta,      
    urlcolor=blue,
    citecolor=blue
}
\makeatletter
\renewcommand*\aa@pageof{, page \thepage{} of \pageref*{LastPage}}
\makeatother
\usepackage[utf8]{inputenc}
\usepackage[switch, modulo]{lineno}
\usepackage{amsmath}
\usepackage{amssymb}
\usepackage{subfigure}
\usepackage{chngcntr}
\usepackage{appendix}
\usepackage{etoolbox}
\usepackage{diagbox}
\usepackage{multicol}
\usepackage{comment}
\usepackage{siunitx}
\usepackage{multirow}
\usepackage{tabulary}
\usepackage[para]{threeparttable}
\usepackage{array,booktabs,longtable,tabularx}
\newcolumntype{L}{>{\raggedright\arraybackslash}X}
\usepackage{ltablex}

\usepackage{euclid}
\usepackage{float} 

\newcommand{\flagship}{\mbox{{\sc \texttt{Flagship}}}\xspace}
\newcommand{\disperse}{\mbox{{\sc \texttt{DisPerSE}}}\xspace}
\newcommand{\trex}{\mbox{{\sc \texttt{T-ReX}}}\xspace}

\begin{document}

\title{Euclid Quick Data Release (Q1)} \subtitle{
 The role of cosmic connectivity in shaping galaxy clusters
}
    
\newcommand{\orcid}[1]{} 
\author{Euclid Collaboration: C.~Gouin\orcid{0000-0002-8837-9953}\thanks{\email{celine.gouin@iap.fr}}\inst{\ref{aff1}}
\and C.~Laigle\orcid{0009-0008-5926-818X}\inst{\ref{aff1}}
\and F.~Sarron\orcid{0000-0001-8376-0360}\inst{\ref{aff2},\ref{aff3}}
\and T.~Bonnaire\orcid{0000-0003-2149-8795}\inst{\ref{aff4}}
\and J.~G.~Sorce\orcid{0000-0002-2307-2432}\inst{\ref{aff5},\ref{aff6}}
\and N.~Aghanim\orcid{0000-0002-6688-8992}\inst{\ref{aff6}}
\and M.~Magliocchetti\orcid{0000-0001-9158-4838}\inst{\ref{aff7}}
\and L.~Quilley\orcid{0009-0008-8375-8605}\inst{\ref{aff8}}
\and P.~Boldrini\inst{\ref{aff1},\ref{aff9}}
\and F.~Durret\orcid{0000-0002-6991-4578}\inst{\ref{aff10}}
\and C.~Pichon\orcid{0000-0003-0695-6735}\inst{\ref{aff1},\ref{aff11}}
\and U.~Kuchner\orcid{0000-0002-0035-5202}\inst{\ref{aff12}}
\and N.~Malavasi\orcid{0000-0001-9033-7958}\inst{\ref{aff13}}
\and K.~Kraljic\orcid{0000-0001-6180-0245}\inst{\ref{aff14}}
\and R.~Gavazzi\orcid{0000-0002-5540-6935}\inst{\ref{aff15},\ref{aff1}}
\and Y.~Kang\orcid{0009-0000-8588-7250}\inst{\ref{aff16}}
\and S.~A.~Stanford\orcid{0000-0003-0122-0841}\inst{\ref{aff17}}
\and P.~Awad\orcid{0000-0002-0428-849X}\inst{\ref{aff18}}
\and B.~Altieri\orcid{0000-0003-3936-0284}\inst{\ref{aff19}}
\and A.~Amara\inst{\ref{aff20}}
\and S.~Andreon\orcid{0000-0002-2041-8784}\inst{\ref{aff21}}
\and N.~Auricchio\orcid{0000-0003-4444-8651}\inst{\ref{aff22}}
\and H.~Aussel\orcid{0000-0002-1371-5705}\inst{\ref{aff23}}
\and C.~Baccigalupi\orcid{0000-0002-8211-1630}\inst{\ref{aff24},\ref{aff25},\ref{aff26},\ref{aff27}}
\and M.~Baldi\orcid{0000-0003-4145-1943}\inst{\ref{aff28},\ref{aff22},\ref{aff29}}
\and A.~Balestra\orcid{0000-0002-6967-261X}\inst{\ref{aff30}}
\and S.~Bardelli\orcid{0000-0002-8900-0298}\inst{\ref{aff22}}
\and A.~Basset\inst{\ref{aff31}}
\and P.~Battaglia\orcid{0000-0002-7337-5909}\inst{\ref{aff22}}
\and F.~Bernardeau\inst{\ref{aff32},\ref{aff1}}
\and A.~Biviano\orcid{0000-0002-0857-0732}\inst{\ref{aff25},\ref{aff24}}
\and A.~Bonchi\orcid{0000-0002-2667-5482}\inst{\ref{aff33}}
\and E.~Branchini\orcid{0000-0002-0808-6908}\inst{\ref{aff34},\ref{aff35},\ref{aff21}}
\and M.~Brescia\orcid{0000-0001-9506-5680}\inst{\ref{aff36},\ref{aff37}}
\and J.~Brinchmann\orcid{0000-0003-4359-8797}\inst{\ref{aff38},\ref{aff39}}
\and S.~Camera\orcid{0000-0003-3399-3574}\inst{\ref{aff40},\ref{aff41},\ref{aff42}}
\and G.~Ca\~nas-Herrera\orcid{0000-0003-2796-2149}\inst{\ref{aff43},\ref{aff44},\ref{aff18}}
\and V.~Capobianco\orcid{0000-0002-3309-7692}\inst{\ref{aff42}}
\and C.~Carbone\orcid{0000-0003-0125-3563}\inst{\ref{aff45}}
\and J.~Carretero\orcid{0000-0002-3130-0204}\inst{\ref{aff46},\ref{aff47}}
\and M.~Castellano\orcid{0000-0001-9875-8263}\inst{\ref{aff48}}
\and G.~Castignani\orcid{0000-0001-6831-0687}\inst{\ref{aff22}}
\and S.~Cavuoti\orcid{0000-0002-3787-4196}\inst{\ref{aff37},\ref{aff49}}
\and K.~C.~Chambers\orcid{0000-0001-6965-7789}\inst{\ref{aff50}}
\and A.~Cimatti\inst{\ref{aff51}}
\and C.~Colodro-Conde\inst{\ref{aff52}}
\and G.~Congedo\orcid{0000-0003-2508-0046}\inst{\ref{aff53}}
\and C.~J.~Conselice\orcid{0000-0003-1949-7638}\inst{\ref{aff54}}
\and L.~Conversi\orcid{0000-0002-6710-8476}\inst{\ref{aff55},\ref{aff19}}
\and Y.~Copin\orcid{0000-0002-5317-7518}\inst{\ref{aff56}}
\and F.~Courbin\orcid{0000-0003-0758-6510}\inst{\ref{aff57},\ref{aff58}}
\and H.~M.~Courtois\orcid{0000-0003-0509-1776}\inst{\ref{aff59}}
\and M.~Cropper\orcid{0000-0003-4571-9468}\inst{\ref{aff60}}
\and A.~Da~Silva\orcid{0000-0002-6385-1609}\inst{\ref{aff61},\ref{aff62}}
\and H.~Degaudenzi\orcid{0000-0002-5887-6799}\inst{\ref{aff16}}
\and S.~de~la~Torre\inst{\ref{aff15}}
\and G.~De~Lucia\orcid{0000-0002-6220-9104}\inst{\ref{aff25}}
\and A.~M.~Di~Giorgio\orcid{0000-0002-4767-2360}\inst{\ref{aff7}}
\and C.~Dolding\orcid{0009-0003-7199-6108}\inst{\ref{aff60}}
\and H.~Dole\orcid{0000-0002-9767-3839}\inst{\ref{aff6}}
\and F.~Dubath\orcid{0000-0002-6533-2810}\inst{\ref{aff16}}
\and C.~A.~J.~Duncan\orcid{0009-0003-3573-0791}\inst{\ref{aff54}}
\and X.~Dupac\inst{\ref{aff19}}
\and A.~Ealet\orcid{0000-0003-3070-014X}\inst{\ref{aff56}}
\and S.~Escoffier\orcid{0000-0002-2847-7498}\inst{\ref{aff63}}
\and M.~Fabricius\orcid{0000-0002-7025-6058}\inst{\ref{aff13},\ref{aff64}}
\and M.~Farina\orcid{0000-0002-3089-7846}\inst{\ref{aff7}}
\and F.~Faustini\orcid{0000-0001-6274-5145}\inst{\ref{aff33},\ref{aff48}}
\and S.~Ferriol\inst{\ref{aff56}}
\and F.~Finelli\orcid{0000-0002-6694-3269}\inst{\ref{aff22},\ref{aff65}}
\and S.~Fotopoulou\orcid{0000-0002-9686-254X}\inst{\ref{aff66}}
\and M.~Frailis\orcid{0000-0002-7400-2135}\inst{\ref{aff25}}
\and E.~Franceschi\orcid{0000-0002-0585-6591}\inst{\ref{aff22}}
\and S.~Galeotta\orcid{0000-0002-3748-5115}\inst{\ref{aff25}}
\and K.~George\orcid{0000-0002-1734-8455}\inst{\ref{aff64}}
\and B.~Gillis\orcid{0000-0002-4478-1270}\inst{\ref{aff53}}
\and C.~Giocoli\orcid{0000-0002-9590-7961}\inst{\ref{aff22},\ref{aff29}}
\and P.~G\'omez-Alvarez\orcid{0000-0002-8594-5358}\inst{\ref{aff67},\ref{aff19}}
\and J.~Gracia-Carpio\inst{\ref{aff13}}
\and B.~R.~Granett\orcid{0000-0003-2694-9284}\inst{\ref{aff21}}
\and A.~Grazian\orcid{0000-0002-5688-0663}\inst{\ref{aff30}}
\and F.~Grupp\inst{\ref{aff13},\ref{aff64}}
\and S.~Gwyn\orcid{0000-0001-8221-8406}\inst{\ref{aff68}}
\and S.~V.~H.~Haugan\orcid{0000-0001-9648-7260}\inst{\ref{aff69}}
\and W.~Holmes\inst{\ref{aff70}}
\and I.~M.~Hook\orcid{0000-0002-2960-978X}\inst{\ref{aff71}}
\and F.~Hormuth\inst{\ref{aff72}}
\and A.~Hornstrup\orcid{0000-0002-3363-0936}\inst{\ref{aff73},\ref{aff74}}
\and P.~Hudelot\inst{\ref{aff1}}
\and K.~Jahnke\orcid{0000-0003-3804-2137}\inst{\ref{aff75}}
\and M.~Jhabvala\inst{\ref{aff76}}
\and B.~Joachimi\orcid{0000-0001-7494-1303}\inst{\ref{aff77}}
\and E.~Keih\"anen\orcid{0000-0003-1804-7715}\inst{\ref{aff78}}
\and S.~Kermiche\orcid{0000-0002-0302-5735}\inst{\ref{aff63}}
\and A.~Kiessling\orcid{0000-0002-2590-1273}\inst{\ref{aff70}}
\and M.~Kilbinger\orcid{0000-0001-9513-7138}\inst{\ref{aff23}}
\and B.~Kubik\orcid{0009-0006-5823-4880}\inst{\ref{aff56}}
\and M.~K\"ummel\orcid{0000-0003-2791-2117}\inst{\ref{aff64}}
\and M.~Kunz\orcid{0000-0002-3052-7394}\inst{\ref{aff79}}
\and H.~Kurki-Suonio\orcid{0000-0002-4618-3063}\inst{\ref{aff80},\ref{aff81}}
\and O.~Lahav\orcid{0000-0002-1134-9035}\inst{\ref{aff77}}
\and Q.~Le~Boulc'h\inst{\ref{aff82}}
\and A.~M.~C.~Le~Brun\orcid{0000-0002-0936-4594}\inst{\ref{aff83}}
\and D.~Le~Mignant\orcid{0000-0002-5339-5515}\inst{\ref{aff15}}
\and P.~Liebing\inst{\ref{aff60}}
\and S.~Ligori\orcid{0000-0003-4172-4606}\inst{\ref{aff42}}
\and P.~B.~Lilje\orcid{0000-0003-4324-7794}\inst{\ref{aff69}}
\and V.~Lindholm\orcid{0000-0003-2317-5471}\inst{\ref{aff80},\ref{aff81}}
\and I.~Lloro\orcid{0000-0001-5966-1434}\inst{\ref{aff84}}
\and G.~Mainetti\orcid{0000-0003-2384-2377}\inst{\ref{aff82}}
\and D.~Maino\inst{\ref{aff85},\ref{aff45},\ref{aff86}}
\and E.~Maiorano\orcid{0000-0003-2593-4355}\inst{\ref{aff22}}
\and O.~Mansutti\orcid{0000-0001-5758-4658}\inst{\ref{aff25}}
\and S.~Marcin\inst{\ref{aff87}}
\and O.~Marggraf\orcid{0000-0001-7242-3852}\inst{\ref{aff88}}
\and M.~Martinelli\orcid{0000-0002-6943-7732}\inst{\ref{aff48},\ref{aff89}}
\and N.~Martinet\orcid{0000-0003-2786-7790}\inst{\ref{aff15}}
\and F.~Marulli\orcid{0000-0002-8850-0303}\inst{\ref{aff90},\ref{aff22},\ref{aff29}}
\and R.~Massey\orcid{0000-0002-6085-3780}\inst{\ref{aff91}}
\and S.~Maurogordato\inst{\ref{aff92}}
\and H.~J.~McCracken\orcid{0000-0002-9489-7765}\inst{\ref{aff1}}
\and E.~Medinaceli\orcid{0000-0002-4040-7783}\inst{\ref{aff22}}
\and S.~Mei\orcid{0000-0002-2849-559X}\inst{\ref{aff93},\ref{aff94}}
\and Y.~Mellier\inst{\ref{aff10},\ref{aff1}}
\and M.~Meneghetti\orcid{0000-0003-1225-7084}\inst{\ref{aff22},\ref{aff29}}
\and E.~Merlin\orcid{0000-0001-6870-8900}\inst{\ref{aff48}}
\and G.~Meylan\inst{\ref{aff95}}
\and A.~Mora\orcid{0000-0002-1922-8529}\inst{\ref{aff96}}
\and M.~Moresco\orcid{0000-0002-7616-7136}\inst{\ref{aff90},\ref{aff22}}
\and L.~Moscardini\orcid{0000-0002-3473-6716}\inst{\ref{aff90},\ref{aff22},\ref{aff29}}
\and R.~Nakajima\orcid{0009-0009-1213-7040}\inst{\ref{aff88}}
\and C.~Neissner\orcid{0000-0001-8524-4968}\inst{\ref{aff97},\ref{aff47}}
\and S.-M.~Niemi\inst{\ref{aff43}}
\and J.~W.~Nightingale\orcid{0000-0002-8987-7401}\inst{\ref{aff98}}
\and C.~Padilla\orcid{0000-0001-7951-0166}\inst{\ref{aff97}}
\and S.~Paltani\orcid{0000-0002-8108-9179}\inst{\ref{aff16}}
\and F.~Pasian\orcid{0000-0002-4869-3227}\inst{\ref{aff25}}
\and J.~A.~Peacock\orcid{0000-0002-1168-8299}\inst{\ref{aff53}}
\and K.~Pedersen\inst{\ref{aff99}}
\and W.~J.~Percival\orcid{0000-0002-0644-5727}\inst{\ref{aff100},\ref{aff101},\ref{aff102}}
\and V.~Pettorino\inst{\ref{aff43}}
\and S.~Pires\orcid{0000-0002-0249-2104}\inst{\ref{aff23}}
\and G.~Polenta\orcid{0000-0003-4067-9196}\inst{\ref{aff33}}
\and M.~Poncet\inst{\ref{aff31}}
\and L.~A.~Popa\inst{\ref{aff103}}
\and L.~Pozzetti\orcid{0000-0001-7085-0412}\inst{\ref{aff22}}
\and F.~Raison\orcid{0000-0002-7819-6918}\inst{\ref{aff13}}
\and R.~Rebolo\orcid{0000-0003-3767-7085}\inst{\ref{aff52},\ref{aff104},\ref{aff105}}
\and A.~Renzi\orcid{0000-0001-9856-1970}\inst{\ref{aff106},\ref{aff107}}
\and J.~Rhodes\orcid{0000-0002-4485-8549}\inst{\ref{aff70}}
\and G.~Riccio\inst{\ref{aff37}}
\and E.~Romelli\orcid{0000-0003-3069-9222}\inst{\ref{aff25}}
\and M.~Roncarelli\orcid{0000-0001-9587-7822}\inst{\ref{aff22}}
\and B.~Rusholme\orcid{0000-0001-7648-4142}\inst{\ref{aff108}}
\and R.~Saglia\orcid{0000-0003-0378-7032}\inst{\ref{aff64},\ref{aff13}}
\and Z.~Sakr\orcid{0000-0002-4823-3757}\inst{\ref{aff109},\ref{aff110},\ref{aff111}}
\and A.~G.~S\'anchez\orcid{0000-0003-1198-831X}\inst{\ref{aff13}}
\and D.~Sapone\orcid{0000-0001-7089-4503}\inst{\ref{aff112}}
\and B.~Sartoris\orcid{0000-0003-1337-5269}\inst{\ref{aff64},\ref{aff25}}
\and J.~A.~Schewtschenko\orcid{0000-0002-4913-6393}\inst{\ref{aff53}}
\and M.~Schirmer\orcid{0000-0003-2568-9994}\inst{\ref{aff75}}
\and P.~Schneider\orcid{0000-0001-8561-2679}\inst{\ref{aff88}}
\and T.~Schrabback\orcid{0000-0002-6987-7834}\inst{\ref{aff113}}
\and M.~Scodeggio\inst{\ref{aff45}}
\and A.~Secroun\orcid{0000-0003-0505-3710}\inst{\ref{aff63}}
\and G.~Seidel\orcid{0000-0003-2907-353X}\inst{\ref{aff75}}
\and S.~Serrano\orcid{0000-0002-0211-2861}\inst{\ref{aff114},\ref{aff115},\ref{aff116}}
\and P.~Simon\inst{\ref{aff88}}
\and C.~Sirignano\orcid{0000-0002-0995-7146}\inst{\ref{aff106},\ref{aff107}}
\and G.~Sirri\orcid{0000-0003-2626-2853}\inst{\ref{aff29}}
\and J.~Skottfelt\orcid{0000-0003-1310-8283}\inst{\ref{aff117}}
\and L.~Stanco\orcid{0000-0002-9706-5104}\inst{\ref{aff107}}
\and J.~Steinwagner\orcid{0000-0001-7443-1047}\inst{\ref{aff13}}
\and P.~Tallada-Cresp\'{i}\orcid{0000-0002-1336-8328}\inst{\ref{aff46},\ref{aff47}}
\and D.~Tavagnacco\orcid{0000-0001-7475-9894}\inst{\ref{aff25}}
\and A.~N.~Taylor\inst{\ref{aff53}}
\and H.~I.~Teplitz\orcid{0000-0002-7064-5424}\inst{\ref{aff118}}
\and I.~Tereno\inst{\ref{aff61},\ref{aff119}}
\and N.~Tessore\orcid{0000-0002-9696-7931}\inst{\ref{aff77}}
\and S.~Toft\orcid{0000-0003-3631-7176}\inst{\ref{aff120},\ref{aff121}}
\and R.~Toledo-Moreo\orcid{0000-0002-2997-4859}\inst{\ref{aff122}}
\and F.~Torradeflot\orcid{0000-0003-1160-1517}\inst{\ref{aff47},\ref{aff46}}
\and A.~Tsyganov\inst{\ref{aff123}}
\and I.~Tutusaus\orcid{0000-0002-3199-0399}\inst{\ref{aff110}}
\and L.~Valenziano\orcid{0000-0002-1170-0104}\inst{\ref{aff22},\ref{aff65}}
\and J.~Valiviita\orcid{0000-0001-6225-3693}\inst{\ref{aff80},\ref{aff81}}
\and T.~Vassallo\orcid{0000-0001-6512-6358}\inst{\ref{aff64},\ref{aff25}}
\and G.~Verdoes~Kleijn\orcid{0000-0001-5803-2580}\inst{\ref{aff124}}
\and A.~Veropalumbo\orcid{0000-0003-2387-1194}\inst{\ref{aff21},\ref{aff35},\ref{aff34}}
\and Y.~Wang\orcid{0000-0002-4749-2984}\inst{\ref{aff118}}
\and J.~Weller\orcid{0000-0002-8282-2010}\inst{\ref{aff64},\ref{aff13}}
\and A.~Zacchei\orcid{0000-0003-0396-1192}\inst{\ref{aff25},\ref{aff24}}
\and G.~Zamorani\orcid{0000-0002-2318-301X}\inst{\ref{aff22}}
\and F.~M.~Zerbi\inst{\ref{aff21}}
\and E.~Zucca\orcid{0000-0002-5845-8132}\inst{\ref{aff22}}
\and V.~Allevato\orcid{0000-0001-7232-5152}\inst{\ref{aff37}}
\and M.~Ballardini\orcid{0000-0003-4481-3559}\inst{\ref{aff125},\ref{aff126},\ref{aff22}}
\and M.~Bolzonella\orcid{0000-0003-3278-4607}\inst{\ref{aff22}}
\and E.~Bozzo\orcid{0000-0002-8201-1525}\inst{\ref{aff16}}
\and C.~Burigana\orcid{0000-0002-3005-5796}\inst{\ref{aff127},\ref{aff65}}
\and R.~Cabanac\orcid{0000-0001-6679-2600}\inst{\ref{aff110}}
\and A.~Cappi\inst{\ref{aff22},\ref{aff92}}
\and D.~Di~Ferdinando\inst{\ref{aff29}}
\and J.~A.~Escartin~Vigo\inst{\ref{aff13}}
\and L.~Gabarra\orcid{0000-0002-8486-8856}\inst{\ref{aff128}}
\and M.~Huertas-Company\orcid{0000-0002-1416-8483}\inst{\ref{aff52},\ref{aff129},\ref{aff130},\ref{aff131}}
\and J.~Mart\'{i}n-Fleitas\orcid{0000-0002-8594-569X}\inst{\ref{aff96}}
\and S.~Matthew\orcid{0000-0001-8448-1697}\inst{\ref{aff53}}
\and M.~Maturi\orcid{0000-0002-3517-2422}\inst{\ref{aff109},\ref{aff132}}
\and N.~Mauri\orcid{0000-0001-8196-1548}\inst{\ref{aff51},\ref{aff29}}
\and R.~B.~Metcalf\orcid{0000-0003-3167-2574}\inst{\ref{aff90},\ref{aff22}}
\and A.~Pezzotta\orcid{0000-0003-0726-2268}\inst{\ref{aff133},\ref{aff13}}
\and M.~P\"ontinen\orcid{0000-0001-5442-2530}\inst{\ref{aff80}}
\and C.~Porciani\orcid{0000-0002-7797-2508}\inst{\ref{aff88}}
\and I.~Risso\orcid{0000-0003-2525-7761}\inst{\ref{aff134}}
\and V.~Scottez\inst{\ref{aff10},\ref{aff135}}
\and M.~Sereno\orcid{0000-0003-0302-0325}\inst{\ref{aff22},\ref{aff29}}
\and M.~Tenti\orcid{0000-0002-4254-5901}\inst{\ref{aff29}}
\and M.~Viel\orcid{0000-0002-2642-5707}\inst{\ref{aff24},\ref{aff25},\ref{aff27},\ref{aff26},\ref{aff136}}
\and M.~Wiesmann\orcid{0009-0000-8199-5860}\inst{\ref{aff69}}
\and Y.~Akrami\orcid{0000-0002-2407-7956}\inst{\ref{aff137},\ref{aff138}}
\and S.~Alvi\orcid{0000-0001-5779-8568}\inst{\ref{aff125}}
\and I.~T.~Andika\orcid{0000-0001-6102-9526}\inst{\ref{aff139},\ref{aff140}}
\and S.~Anselmi\orcid{0000-0002-3579-9583}\inst{\ref{aff107},\ref{aff106},\ref{aff141}}
\and M.~Archidiacono\orcid{0000-0003-4952-9012}\inst{\ref{aff85},\ref{aff86}}
\and F.~Atrio-Barandela\orcid{0000-0002-2130-2513}\inst{\ref{aff142}}
\and A.~Balaguera-Antolinez\orcid{0000-0001-5028-3035}\inst{\ref{aff52}}
\and C.~Benoist\inst{\ref{aff92}}
\and K.~Benson\inst{\ref{aff60}}
\and P.~Bergamini\orcid{0000-0003-1383-9414}\inst{\ref{aff85},\ref{aff22}}
\and D.~Bertacca\orcid{0000-0002-2490-7139}\inst{\ref{aff106},\ref{aff30},\ref{aff107}}
\and M.~Bethermin\orcid{0000-0002-3915-2015}\inst{\ref{aff14}}
\and A.~Blanchard\orcid{0000-0001-8555-9003}\inst{\ref{aff110}}
\and L.~Blot\orcid{0000-0002-9622-7167}\inst{\ref{aff143},\ref{aff141}}
\and H.~B\"ohringer\orcid{0000-0001-8241-4204}\inst{\ref{aff13},\ref{aff144},\ref{aff145}}
\and M.~L.~Brown\orcid{0000-0002-0370-8077}\inst{\ref{aff54}}
\and S.~Bruton\orcid{0000-0002-6503-5218}\inst{\ref{aff146}}
\and A.~Calabro\orcid{0000-0003-2536-1614}\inst{\ref{aff48}}
\and B.~Camacho~Quevedo\orcid{0000-0002-8789-4232}\inst{\ref{aff114},\ref{aff116}}
\and F.~Caro\inst{\ref{aff48}}
\and C.~S.~Carvalho\inst{\ref{aff119}}
\and T.~Castro\orcid{0000-0002-6292-3228}\inst{\ref{aff25},\ref{aff26},\ref{aff24},\ref{aff136}}
\and F.~Cogato\orcid{0000-0003-4632-6113}\inst{\ref{aff90},\ref{aff22}}
\and A.~R.~Cooray\orcid{0000-0002-3892-0190}\inst{\ref{aff147}}
\and O.~Cucciati\orcid{0000-0002-9336-7551}\inst{\ref{aff22}}
\and S.~Davini\orcid{0000-0003-3269-1718}\inst{\ref{aff35}}
\and F.~De~Paolis\orcid{0000-0001-6460-7563}\inst{\ref{aff148},\ref{aff149},\ref{aff150}}
\and G.~Desprez\orcid{0000-0001-8325-1742}\inst{\ref{aff124}}
\and A.~D\'iaz-S\'anchez\orcid{0000-0003-0748-4768}\inst{\ref{aff151}}
\and J.~J.~Diaz\inst{\ref{aff129}}
\and S.~Di~Domizio\orcid{0000-0003-2863-5895}\inst{\ref{aff34},\ref{aff35}}
\and J.~M.~Diego\orcid{0000-0001-9065-3926}\inst{\ref{aff152}}
\and P.-A.~Duc\orcid{0000-0003-3343-6284}\inst{\ref{aff14}}
\and A.~Enia\orcid{0000-0002-0200-2857}\inst{\ref{aff28},\ref{aff22}}
\and Y.~Fang\inst{\ref{aff64}}
\and A.~G.~Ferrari\orcid{0009-0005-5266-4110}\inst{\ref{aff29}}
\and P.~G.~Ferreira\orcid{0000-0002-3021-2851}\inst{\ref{aff128}}
\and A.~Finoguenov\orcid{0000-0002-4606-5403}\inst{\ref{aff80}}
\and A.~Fontana\orcid{0000-0003-3820-2823}\inst{\ref{aff48}}
\and A.~Franco\orcid{0000-0002-4761-366X}\inst{\ref{aff149},\ref{aff148},\ref{aff150}}
\and K.~Ganga\orcid{0000-0001-8159-8208}\inst{\ref{aff93}}
\and J.~Garc\'ia-Bellido\orcid{0000-0002-9370-8360}\inst{\ref{aff137}}
\and T.~Gasparetto\orcid{0000-0002-7913-4866}\inst{\ref{aff25}}
\and V.~Gautard\inst{\ref{aff153}}
\and E.~Gaztanaga\orcid{0000-0001-9632-0815}\inst{\ref{aff116},\ref{aff114},\ref{aff154}}
\and F.~Giacomini\orcid{0000-0002-3129-2814}\inst{\ref{aff29}}
\and F.~Gianotti\orcid{0000-0003-4666-119X}\inst{\ref{aff22}}
\and G.~Gozaliasl\orcid{0000-0002-0236-919X}\inst{\ref{aff155},\ref{aff80}}
\and M.~Guidi\orcid{0000-0001-9408-1101}\inst{\ref{aff28},\ref{aff22}}
\and C.~M.~Gutierrez\orcid{0000-0001-7854-783X}\inst{\ref{aff156}}
\and A.~Hall\orcid{0000-0002-3139-8651}\inst{\ref{aff53}}
\and W.~G.~Hartley\inst{\ref{aff16}}
\and S.~Hemmati\orcid{0000-0003-2226-5395}\inst{\ref{aff108}}
\and C.~Hern\'andez-Monteagudo\orcid{0000-0001-5471-9166}\inst{\ref{aff105},\ref{aff52}}
\and H.~Hildebrandt\orcid{0000-0002-9814-3338}\inst{\ref{aff157}}
\and J.~Hjorth\orcid{0000-0002-4571-2306}\inst{\ref{aff99}}
\and J.~J.~E.~Kajava\orcid{0000-0002-3010-8333}\inst{\ref{aff158},\ref{aff159}}
\and V.~Kansal\orcid{0000-0002-4008-6078}\inst{\ref{aff160},\ref{aff161}}
\and D.~Karagiannis\orcid{0000-0002-4927-0816}\inst{\ref{aff125},\ref{aff162}}
\and K.~Kiiveri\inst{\ref{aff78}}
\and C.~C.~Kirkpatrick\inst{\ref{aff78}}
\and S.~Kruk\orcid{0000-0001-8010-8879}\inst{\ref{aff19}}
\and J.~Le~Graet\orcid{0000-0001-6523-7971}\inst{\ref{aff63}}
\and L.~Legrand\orcid{0000-0003-0610-5252}\inst{\ref{aff163},\ref{aff164}}
\and M.~Lembo\orcid{0000-0002-5271-5070}\inst{\ref{aff125},\ref{aff126}}
\and F.~Lepori\orcid{0009-0000-5061-7138}\inst{\ref{aff165}}
\and G.~Leroy\orcid{0009-0004-2523-4425}\inst{\ref{aff166},\ref{aff91}}
\and G.~F.~Lesci\orcid{0000-0002-4607-2830}\inst{\ref{aff90},\ref{aff22}}
\and J.~Lesgourgues\orcid{0000-0001-7627-353X}\inst{\ref{aff167}}
\and L.~Leuzzi\orcid{0009-0006-4479-7017}\inst{\ref{aff90},\ref{aff22}}
\and T.~I.~Liaudat\orcid{0000-0002-9104-314X}\inst{\ref{aff168}}
\and A.~Loureiro\orcid{0000-0002-4371-0876}\inst{\ref{aff169},\ref{aff170}}
\and J.~Macias-Perez\orcid{0000-0002-5385-2763}\inst{\ref{aff171}}
\and G.~Maggio\orcid{0000-0003-4020-4836}\inst{\ref{aff25}}
\and E.~A.~Magnier\orcid{0000-0002-7965-2815}\inst{\ref{aff50}}
\and F.~Mannucci\orcid{0000-0002-4803-2381}\inst{\ref{aff172}}
\and R.~Maoli\orcid{0000-0002-6065-3025}\inst{\ref{aff173},\ref{aff48}}
\and C.~J.~A.~P.~Martins\orcid{0000-0002-4886-9261}\inst{\ref{aff174},\ref{aff38}}
\and L.~Maurin\orcid{0000-0002-8406-0857}\inst{\ref{aff6}}
\and M.~Migliaccio\inst{\ref{aff175},\ref{aff176}}
\and M.~Miluzio\inst{\ref{aff19},\ref{aff177}}
\and P.~Monaco\orcid{0000-0003-2083-7564}\inst{\ref{aff178},\ref{aff25},\ref{aff26},\ref{aff24}}
\and A.~Montoro\orcid{0000-0003-4730-8590}\inst{\ref{aff116},\ref{aff114}}
\and C.~Moretti\orcid{0000-0003-3314-8936}\inst{\ref{aff27},\ref{aff136},\ref{aff25},\ref{aff24},\ref{aff26}}
\and G.~Morgante\inst{\ref{aff22}}
\and S.~Nadathur\orcid{0000-0001-9070-3102}\inst{\ref{aff154}}
\and K.~Naidoo\orcid{0000-0002-9182-1802}\inst{\ref{aff154}}
\and A.~Navarro-Alsina\orcid{0000-0002-3173-2592}\inst{\ref{aff88}}
\and S.~Nesseris\orcid{0000-0002-0567-0324}\inst{\ref{aff137}}
\and F.~Passalacqua\orcid{0000-0002-8606-4093}\inst{\ref{aff106},\ref{aff107}}
\and K.~Paterson\orcid{0000-0001-8340-3486}\inst{\ref{aff75}}
\and L.~Patrizii\inst{\ref{aff29}}
\and A.~Pisani\orcid{0000-0002-6146-4437}\inst{\ref{aff63},\ref{aff179}}
\and D.~Potter\orcid{0000-0002-0757-5195}\inst{\ref{aff165}}
\and S.~Quai\orcid{0000-0002-0449-8163}\inst{\ref{aff90},\ref{aff22}}
\and M.~Radovich\orcid{0000-0002-3585-866X}\inst{\ref{aff30}}
\and S.~Sacquegna\orcid{0000-0002-8433-6630}\inst{\ref{aff148},\ref{aff149},\ref{aff150}}
\and M.~Sahl\'en\orcid{0000-0003-0973-4804}\inst{\ref{aff180}}
\and D.~B.~Sanders\orcid{0000-0002-1233-9998}\inst{\ref{aff50}}
\and E.~Sarpa\orcid{0000-0002-1256-655X}\inst{\ref{aff27},\ref{aff136},\ref{aff26}}
\and A.~Schneider\orcid{0000-0001-7055-8104}\inst{\ref{aff165}}
\and D.~Sciotti\orcid{0009-0008-4519-2620}\inst{\ref{aff48},\ref{aff89}}
\and D.~Scognamiglio\orcid{0000-0001-8450-7885}\inst{\ref{aff70}}
\and E.~Sellentin\inst{\ref{aff181},\ref{aff18}}
\and F.~Shankar\orcid{0000-0001-8973-5051}\inst{\ref{aff182}}
\and L.~C.~Smith\orcid{0000-0002-3259-2771}\inst{\ref{aff183}}
\and K.~Tanidis\orcid{0000-0001-9843-5130}\inst{\ref{aff128}}
\and G.~Testera\inst{\ref{aff35}}
\and R.~Teyssier\orcid{0000-0001-7689-0933}\inst{\ref{aff179}}
\and S.~Tosi\orcid{0000-0002-7275-9193}\inst{\ref{aff34},\ref{aff35},\ref{aff21}}
\and A.~Troja\orcid{0000-0003-0239-4595}\inst{\ref{aff106},\ref{aff107}}
\and M.~Tucci\inst{\ref{aff16}}
\and C.~Valieri\inst{\ref{aff29}}
\and A.~Venhola\orcid{0000-0001-6071-4564}\inst{\ref{aff184}}
\and D.~Vergani\orcid{0000-0003-0898-2216}\inst{\ref{aff22}}
\and G.~Verza\orcid{0000-0002-1886-8348}\inst{\ref{aff185}}
\and P.~Vielzeuf\orcid{0000-0003-2035-9339}\inst{\ref{aff63}}
\and N.~A.~Walton\orcid{0000-0003-3983-8778}\inst{\ref{aff183}}
\and D.~Scott\orcid{0000-0002-6878-9840}\inst{\ref{aff186}}}
										   
\institute{Institut d'Astrophysique de Paris, UMR 7095, CNRS, and Sorbonne Universit\'e, 98 bis boulevard Arago, 75014 Paris, France\label{aff1}
\and
Institut de Recherche en Informatique de Toulouse (IRIT), Universit\'e de Toulouse, CNRS, Toulouse INP, UT3, 31062 Toulouse, France\label{aff2}
\and
Laboratoire MCD, Centre de Biologie Int\'egrative (CBI), Universit\'e de Toulouse, CNRS, UT3, 31062 Toulouse, France\label{aff3}
\and
Laboratoire de Physique de l'\'Ecole Normale Sup\'erieure, ENS, Universit\'e PSL, CNRS, Sorbonne Universit\'e, 75005 Paris, France\label{aff4}
\and
Univ. Lille, CNRS, Centrale Lille, UMR 9189 CRIStAL, 59000 Lille, France\label{aff5}
\and
Universit\'e Paris-Saclay, CNRS, Institut d'astrophysique spatiale, 91405, Orsay, France\label{aff6}
\and
INAF-Istituto di Astrofisica e Planetologia Spaziali, via del Fosso del Cavaliere, 100, 00100 Roma, Italy\label{aff7}
\and
Centre de Recherche Astrophysique de Lyon, UMR5574, CNRS, Universit\'e Claude Bernard Lyon 1, ENS de Lyon, 69230, Saint-Genis-Laval, France\label{aff8}
\and
Observatoire de Paris, PSL Research University 61, avenue de l'Observatoire, 75014 Paris, France\label{aff9}
\and
Institut d'Astrophysique de Paris, 98bis Boulevard Arago, 75014, Paris, France\label{aff10}
\and
Kyung Hee University, Dept. of Astronomy \& Space Science, Yongin-shi, Gyeonggi-do 17104, Republic of Korea\label{aff11}
\and
School of Physics and Astronomy, University of Nottingham, University Park, Nottingham NG7 2RD, UK\label{aff12}
\and
Max Planck Institute for Extraterrestrial Physics, Giessenbachstr. 1, 85748 Garching, Germany\label{aff13}
\and
Universit\'e de Strasbourg, CNRS, Observatoire astronomique de Strasbourg, UMR 7550, 67000 Strasbourg, France\label{aff14}
\and
Aix-Marseille Universit\'e, CNRS, CNES, LAM, Marseille, France\label{aff15}
\and
Department of Astronomy, University of Geneva, ch. d'Ecogia 16, 1290 Versoix, Switzerland\label{aff16}
\and
Department of Physics and Astronomy, University of California, Davis, CA 95616, USA\label{aff17}
\and
Leiden Observatory, Leiden University, Einsteinweg 55, 2333 CC Leiden, The Netherlands\label{aff18}
\and
ESAC/ESA, Camino Bajo del Castillo, s/n., Urb. Villafranca del Castillo, 28692 Villanueva de la Ca\~nada, Madrid, Spain\label{aff19}
\and
School of Mathematics and Physics, University of Surrey, Guildford, Surrey, GU2 7XH, UK\label{aff20}
\and
INAF-Osservatorio Astronomico di Brera, Via Brera 28, 20122 Milano, Italy\label{aff21}
\and
INAF-Osservatorio di Astrofisica e Scienza dello Spazio di Bologna, Via Piero Gobetti 93/3, 40129 Bologna, Italy\label{aff22}
\and
Universit\'e Paris-Saclay, Universit\'e Paris Cit\'e, CEA, CNRS, AIM, 91191, Gif-sur-Yvette, France\label{aff23}
\and
IFPU, Institute for Fundamental Physics of the Universe, via Beirut 2, 34151 Trieste, Italy\label{aff24}
\and
INAF-Osservatorio Astronomico di Trieste, Via G. B. Tiepolo 11, 34143 Trieste, Italy\label{aff25}
\and
INFN, Sezione di Trieste, Via Valerio 2, 34127 Trieste TS, Italy\label{aff26}
\and
SISSA, International School for Advanced Studies, Via Bonomea 265, 34136 Trieste TS, Italy\label{aff27}
\and
Dipartimento di Fisica e Astronomia, Universit\`a di Bologna, Via Gobetti 93/2, 40129 Bologna, Italy\label{aff28}
\and
INFN-Sezione di Bologna, Viale Berti Pichat 6/2, 40127 Bologna, Italy\label{aff29}
\and
INAF-Osservatorio Astronomico di Padova, Via dell'Osservatorio 5, 35122 Padova, Italy\label{aff30}
\and
Centre National d'Etudes Spatiales -- Centre spatial de Toulouse, 18 avenue Edouard Belin, 31401 Toulouse Cedex 9, France\label{aff31}
\and
Institut de Physique Th\'eorique, CEA, CNRS, Universit\'e Paris-Saclay 91191 Gif-sur-Yvette Cedex, France\label{aff32}
\and
Space Science Data Center, Italian Space Agency, via del Politecnico snc, 00133 Roma, Italy\label{aff33}
\and
Dipartimento di Fisica, Universit\`a di Genova, Via Dodecaneso 33, 16146, Genova, Italy\label{aff34}
\and
INFN-Sezione di Genova, Via Dodecaneso 33, 16146, Genova, Italy\label{aff35}
\and
Department of Physics "E. Pancini", University Federico II, Via Cinthia 6, 80126, Napoli, Italy\label{aff36}
\and
INAF-Osservatorio Astronomico di Capodimonte, Via Moiariello 16, 80131 Napoli, Italy\label{aff37}
\and
Instituto de Astrof\'isica e Ci\^encias do Espa\c{c}o, Universidade do Porto, CAUP, Rua das Estrelas, PT4150-762 Porto, Portugal\label{aff38}
\and
Faculdade de Ci\^encias da Universidade do Porto, Rua do Campo de Alegre, 4150-007 Porto, Portugal\label{aff39}
\and
Dipartimento di Fisica, Universit\`a degli Studi di Torino, Via P. Giuria 1, 10125 Torino, Italy\label{aff40}
\and
INFN-Sezione di Torino, Via P. Giuria 1, 10125 Torino, Italy\label{aff41}
\and
INAF-Osservatorio Astrofisico di Torino, Via Osservatorio 20, 10025 Pino Torinese (TO), Italy\label{aff42}
\and
European Space Agency/ESTEC, Keplerlaan 1, 2201 AZ Noordwijk, The Netherlands\label{aff43}
\and
Institute Lorentz, Leiden University, Niels Bohrweg 2, 2333 CA Leiden, The Netherlands\label{aff44}
\and
INAF-IASF Milano, Via Alfonso Corti 12, 20133 Milano, Italy\label{aff45}
\and
Centro de Investigaciones Energ\'eticas, Medioambientales y Tecnol\'ogicas (CIEMAT), Avenida Complutense 40, 28040 Madrid, Spain\label{aff46}
\and
Port d'Informaci\'{o} Cient\'{i}fica, Campus UAB, C. Albareda s/n, 08193 Bellaterra (Barcelona), Spain\label{aff47}
\and
INAF-Osservatorio Astronomico di Roma, Via Frascati 33, 00078 Monteporzio Catone, Italy\label{aff48}
\and
INFN section of Naples, Via Cinthia 6, 80126, Napoli, Italy\label{aff49}
\and
Institute for Astronomy, University of Hawaii, 2680 Woodlawn Drive, Honolulu, HI 96822, USA\label{aff50}
\and
Dipartimento di Fisica e Astronomia "Augusto Righi" - Alma Mater Studiorum Universit\`a di Bologna, Viale Berti Pichat 6/2, 40127 Bologna, Italy\label{aff51}
\and
Instituto de Astrof\'{\i}sica de Canarias, V\'{\i}a L\'actea, 38205 La Laguna, Tenerife, Spain\label{aff52}
\and
Institute for Astronomy, University of Edinburgh, Royal Observatory, Blackford Hill, Edinburgh EH9 3HJ, UK\label{aff53}
\and
Jodrell Bank Centre for Astrophysics, Department of Physics and Astronomy, University of Manchester, Oxford Road, Manchester M13 9PL, UK\label{aff54}
\and
European Space Agency/ESRIN, Largo Galileo Galilei 1, 00044 Frascati, Roma, Italy\label{aff55}
\and
Universit\'e Claude Bernard Lyon 1, CNRS/IN2P3, IP2I Lyon, UMR 5822, Villeurbanne, F-69100, France\label{aff56}
\and
Institut de Ci\`{e}ncies del Cosmos (ICCUB), Universitat de Barcelona (IEEC-UB), Mart\'{i} i Franqu\`{e}s 1, 08028 Barcelona, Spain\label{aff57}
\and
Instituci\'o Catalana de Recerca i Estudis Avan\c{c}ats (ICREA), Passeig de Llu\'{\i}s Companys 23, 08010 Barcelona, Spain\label{aff58}
\and
UCB Lyon 1, CNRS/IN2P3, IUF, IP2I Lyon, 4 rue Enrico Fermi, 69622 Villeurbanne, France\label{aff59}
\and
Mullard Space Science Laboratory, University College London, Holmbury St Mary, Dorking, Surrey RH5 6NT, UK\label{aff60}
\and
Departamento de F\'isica, Faculdade de Ci\^encias, Universidade de Lisboa, Edif\'icio C8, Campo Grande, PT1749-016 Lisboa, Portugal\label{aff61}
\and
Instituto de Astrof\'isica e Ci\^encias do Espa\c{c}o, Faculdade de Ci\^encias, Universidade de Lisboa, Campo Grande, 1749-016 Lisboa, Portugal\label{aff62}
\and
Aix-Marseille Universit\'e, CNRS/IN2P3, CPPM, Marseille, France\label{aff63}
\and
Universit\"ats-Sternwarte M\"unchen, Fakult\"at f\"ur Physik, Ludwig-Maximilians-Universit\"at M\"unchen, Scheinerstrasse 1, 81679 M\"unchen, Germany\label{aff64}
\and
INFN-Bologna, Via Irnerio 46, 40126 Bologna, Italy\label{aff65}
\and
School of Physics, HH Wills Physics Laboratory, University of Bristol, Tyndall Avenue, Bristol, BS8 1TL, UK\label{aff66}
\and
FRACTAL S.L.N.E., calle Tulip\'an 2, Portal 13 1A, 28231, Las Rozas de Madrid, Spain\label{aff67}
\and
NRC Herzberg, 5071 West Saanich Rd, Victoria, BC V9E 2E7, Canada\label{aff68}
\and
Institute of Theoretical Astrophysics, University of Oslo, P.O. Box 1029 Blindern, 0315 Oslo, Norway\label{aff69}
\and
Jet Propulsion Laboratory, California Institute of Technology, 4800 Oak Grove Drive, Pasadena, CA, 91109, USA\label{aff70}
\and
Department of Physics, Lancaster University, Lancaster, LA1 4YB, UK\label{aff71}
\and
Felix Hormuth Engineering, Goethestr. 17, 69181 Leimen, Germany\label{aff72}
\and
Technical University of Denmark, Elektrovej 327, 2800 Kgs. Lyngby, Denmark\label{aff73}
\and
Cosmic Dawn Center (DAWN), Denmark\label{aff74}
\and
Max-Planck-Institut f\"ur Astronomie, K\"onigstuhl 17, 69117 Heidelberg, Germany\label{aff75}
\and
NASA Goddard Space Flight Center, Greenbelt, MD 20771, USA\label{aff76}
\and
Department of Physics and Astronomy, University College London, Gower Street, London WC1E 6BT, UK\label{aff77}
\and
Department of Physics and Helsinki Institute of Physics, Gustaf H\"allstr\"omin katu 2, 00014 University of Helsinki, Finland\label{aff78}
\and
Universit\'e de Gen\`eve, D\'epartement de Physique Th\'eorique and Centre for Astroparticle Physics, 24 quai Ernest-Ansermet, CH-1211 Gen\`eve 4, Switzerland\label{aff79}
\and
Department of Physics, P.O. Box 64, 00014 University of Helsinki, Finland\label{aff80}
\and
Helsinki Institute of Physics, Gustaf H{\"a}llstr{\"o}min katu 2, University of Helsinki, Helsinki, Finland\label{aff81}
\and
Centre de Calcul de l'IN2P3/CNRS, 21 avenue Pierre de Coubertin 69627 Villeurbanne Cedex, France\label{aff82}
\and
Laboratoire d'etude de l'Univers et des phenomenes eXtremes, Observatoire de Paris, Universit\'e PSL, Sorbonne Universit\'e, CNRS, 92190 Meudon, France\label{aff83}
\and
SKA Observatory, Jodrell Bank, Lower Withington, Macclesfield, Cheshire SK11 9FT, UK\label{aff84}
\and
Dipartimento di Fisica "Aldo Pontremoli", Universit\`a degli Studi di Milano, Via Celoria 16, 20133 Milano, Italy\label{aff85}
\and
INFN-Sezione di Milano, Via Celoria 16, 20133 Milano, Italy\label{aff86}
\and
University of Applied Sciences and Arts of Northwestern Switzerland, School of Computer Science, 5210 Windisch, Switzerland\label{aff87}
\and
Universit\"at Bonn, Argelander-Institut f\"ur Astronomie, Auf dem H\"ugel 71, 53121 Bonn, Germany\label{aff88}
\and
INFN-Sezione di Roma, Piazzale Aldo Moro, 2 - c/o Dipartimento di Fisica, Edificio G. Marconi, 00185 Roma, Italy\label{aff89}
\and
Dipartimento di Fisica e Astronomia "Augusto Righi" - Alma Mater Studiorum Universit\`a di Bologna, via Piero Gobetti 93/2, 40129 Bologna, Italy\label{aff90}
\and
Department of Physics, Institute for Computational Cosmology, Durham University, South Road, Durham, DH1 3LE, UK\label{aff91}
\and
Universit\'e C\^{o}te d'Azur, Observatoire de la C\^{o}te d'Azur, CNRS, Laboratoire Lagrange, Bd de l'Observatoire, CS 34229, 06304 Nice cedex 4, France\label{aff92}
\and
Universit\'e Paris Cit\'e, CNRS, Astroparticule et Cosmologie, 75013 Paris, France\label{aff93}
\and
CNRS-UCB International Research Laboratory, Centre Pierre Binetruy, IRL2007, CPB-IN2P3, Berkeley, USA\label{aff94}
\and
Institute of Physics, Laboratory of Astrophysics, Ecole Polytechnique F\'ed\'erale de Lausanne (EPFL), Observatoire de Sauverny, 1290 Versoix, Switzerland\label{aff95}
\and
Aurora Technology for European Space Agency (ESA), Camino bajo del Castillo, s/n, Urbanizacion Villafranca del Castillo, Villanueva de la Ca\~nada, 28692 Madrid, Spain\label{aff96}
\and
Institut de F\'{i}sica d'Altes Energies (IFAE), The Barcelona Institute of Science and Technology, Campus UAB, 08193 Bellaterra (Barcelona), Spain\label{aff97}
\and
School of Mathematics, Statistics and Physics, Newcastle University, Herschel Building, Newcastle-upon-Tyne, NE1 7RU, UK\label{aff98}
\and
DARK, Niels Bohr Institute, University of Copenhagen, Jagtvej 155, 2200 Copenhagen, Denmark\label{aff99}
\and
Waterloo Centre for Astrophysics, University of Waterloo, Waterloo, Ontario N2L 3G1, Canada\label{aff100}
\and
Department of Physics and Astronomy, University of Waterloo, Waterloo, Ontario N2L 3G1, Canada\label{aff101}
\and
Perimeter Institute for Theoretical Physics, Waterloo, Ontario N2L 2Y5, Canada\label{aff102}
\and
Institute of Space Science, Str. Atomistilor, nr. 409 M\u{a}gurele, Ilfov, 077125, Romania\label{aff103}
\and
Consejo Superior de Investigaciones Cientificas, Calle Serrano 117, 28006 Madrid, Spain\label{aff104}
\and
Universidad de La Laguna, Departamento de Astrof\'{\i}sica, 38206 La Laguna, Tenerife, Spain\label{aff105}
\and
Dipartimento di Fisica e Astronomia "G. Galilei", Universit\`a di Padova, Via Marzolo 8, 35131 Padova, Italy\label{aff106}
\and
INFN-Padova, Via Marzolo 8, 35131 Padova, Italy\label{aff107}
\and
Caltech/IPAC, 1200 E. California Blvd., Pasadena, CA 91125, USA\label{aff108}
\and
Institut f\"ur Theoretische Physik, University of Heidelberg, Philosophenweg 16, 69120 Heidelberg, Germany\label{aff109}
\and
Institut de Recherche en Astrophysique et Plan\'etologie (IRAP), Universit\'e de Toulouse, CNRS, UPS, CNES, 14 Av. Edouard Belin, 31400 Toulouse, France\label{aff110}
\and
Universit\'e St Joseph; Faculty of Sciences, Beirut, Lebanon\label{aff111}
\and
Departamento de F\'isica, FCFM, Universidad de Chile, Blanco Encalada 2008, Santiago, Chile\label{aff112}
\and
Universit\"at Innsbruck, Institut f\"ur Astro- und Teilchenphysik, Technikerstr. 25/8, 6020 Innsbruck, Austria\label{aff113}
\and
Institut d'Estudis Espacials de Catalunya (IEEC),  Edifici RDIT, Campus UPC, 08860 Castelldefels, Barcelona, Spain\label{aff114}
\and
Satlantis, University Science Park, Sede Bld 48940, Leioa-Bilbao, Spain\label{aff115}
\and
Institute of Space Sciences (ICE, CSIC), Campus UAB, Carrer de Can Magrans, s/n, 08193 Barcelona, Spain\label{aff116}
\and
Centre for Electronic Imaging, Open University, Walton Hall, Milton Keynes, MK7~6AA, UK\label{aff117}
\and
Infrared Processing and Analysis Center, California Institute of Technology, Pasadena, CA 91125, USA\label{aff118}
\and
Instituto de Astrof\'isica e Ci\^encias do Espa\c{c}o, Faculdade de Ci\^encias, Universidade de Lisboa, Tapada da Ajuda, 1349-018 Lisboa, Portugal\label{aff119}
\and
Cosmic Dawn Center (DAWN)\label{aff120}
\and
Niels Bohr Institute, University of Copenhagen, Jagtvej 128, 2200 Copenhagen, Denmark\label{aff121}
\and
Universidad Polit\'ecnica de Cartagena, Departamento de Electr\'onica y Tecnolog\'ia de Computadoras,  Plaza del Hospital 1, 30202 Cartagena, Spain\label{aff122}
\and
Centre for Information Technology, University of Groningen, P.O. Box 11044, 9700 CA Groningen, The Netherlands\label{aff123}
\and
Kapteyn Astronomical Institute, University of Groningen, PO Box 800, 9700 AV Groningen, The Netherlands\label{aff124}
\and
Dipartimento di Fisica e Scienze della Terra, Universit\`a degli Studi di Ferrara, Via Giuseppe Saragat 1, 44122 Ferrara, Italy\label{aff125}
\and
Istituto Nazionale di Fisica Nucleare, Sezione di Ferrara, Via Giuseppe Saragat 1, 44122 Ferrara, Italy\label{aff126}
\and
INAF, Istituto di Radioastronomia, Via Piero Gobetti 101, 40129 Bologna, Italy\label{aff127}
\and
Department of Physics, Oxford University, Keble Road, Oxford OX1 3RH, UK\label{aff128}
\and
Instituto de Astrof\'isica de Canarias (IAC); Departamento de Astrof\'isica, Universidad de La Laguna (ULL), 38200, La Laguna, Tenerife, Spain\label{aff129}
\and
Universit\'e PSL, Observatoire de Paris, Sorbonne Universit\'e, CNRS, LERMA, 75014, Paris, France\label{aff130}
\and
Universit\'e Paris-Cit\'e, 5 Rue Thomas Mann, 75013, Paris, France\label{aff131}
\and
Zentrum f\"ur Astronomie, Universit\"at Heidelberg, Philosophenweg 12, 69120 Heidelberg, Germany\label{aff132}
\and
INAF - Osservatorio Astronomico di Brera, via Emilio Bianchi 46, 23807 Merate, Italy\label{aff133}
\and
INAF-Osservatorio Astronomico di Brera, Via Brera 28, 20122 Milano, Italy, and INFN-Sezione di Genova, Via Dodecaneso 33, 16146, Genova, Italy\label{aff134}
\and
ICL, Junia, Universit\'e Catholique de Lille, LITL, 59000 Lille, France\label{aff135}
\and
ICSC - Centro Nazionale di Ricerca in High Performance Computing, Big Data e Quantum Computing, Via Magnanelli 2, Bologna, Italy\label{aff136}
\and
Instituto de F\'isica Te\'orica UAM-CSIC, Campus de Cantoblanco, 28049 Madrid, Spain\label{aff137}
\and
CERCA/ISO, Department of Physics, Case Western Reserve University, 10900 Euclid Avenue, Cleveland, OH 44106, USA\label{aff138}
\and
Technical University of Munich, TUM School of Natural Sciences, Physics Department, James-Franck-Str.~1, 85748 Garching, Germany\label{aff139}
\and
Max-Planck-Institut f\"ur Astrophysik, Karl-Schwarzschild-Str.~1, 85748 Garching, Germany\label{aff140}
\and
Laboratoire Univers et Th\'eorie, Observatoire de Paris, Universit\'e PSL, Universit\'e Paris Cit\'e, CNRS, 92190 Meudon, France\label{aff141}
\and
Departamento de F{\'\i}sica Fundamental. Universidad de Salamanca. Plaza de la Merced s/n. 37008 Salamanca, Spain\label{aff142}
\and
Center for Data-Driven Discovery, Kavli IPMU (WPI), UTIAS, The University of Tokyo, Kashiwa, Chiba 277-8583, Japan\label{aff143}
\and
Ludwig-Maximilians-University, Schellingstrasse 4, 80799 Munich, Germany\label{aff144}
\and
Max-Planck-Institut f\"ur Physik, Boltzmannstr. 8, 85748 Garching, Germany\label{aff145}
\and
California Institute of Technology, 1200 E California Blvd, Pasadena, CA 91125, USA\label{aff146}
\and
Department of Physics \& Astronomy, University of California Irvine, Irvine CA 92697, USA\label{aff147}
\and
Department of Mathematics and Physics E. De Giorgi, University of Salento, Via per Arnesano, CP-I93, 73100, Lecce, Italy\label{aff148}
\and
INFN, Sezione di Lecce, Via per Arnesano, CP-193, 73100, Lecce, Italy\label{aff149}
\and
INAF-Sezione di Lecce, c/o Dipartimento Matematica e Fisica, Via per Arnesano, 73100, Lecce, Italy\label{aff150}
\and
Departamento F\'isica Aplicada, Universidad Polit\'ecnica de Cartagena, Campus Muralla del Mar, 30202 Cartagena, Murcia, Spain\label{aff151}
\and
Instituto de F\'isica de Cantabria, Edificio Juan Jord\'a, Avenida de los Castros, 39005 Santander, Spain\label{aff152}
\and
CEA Saclay, DFR/IRFU, Service d'Astrophysique, Bat. 709, 91191 Gif-sur-Yvette, France\label{aff153}
\and
Institute of Cosmology and Gravitation, University of Portsmouth, Portsmouth PO1 3FX, UK\label{aff154}
\and
Department of Computer Science, Aalto University, PO Box 15400, Espoo, FI-00 076, Finland\label{aff155}
\and
Instituto de Astrof\'\i sica de Canarias, c/ Via Lactea s/n, La Laguna 38200, Spain. Departamento de Astrof\'\i sica de la Universidad de La Laguna, Avda. Francisco Sanchez, La Laguna, 38200, Spain\label{aff156}
\and
Ruhr University Bochum, Faculty of Physics and Astronomy, Astronomical Institute (AIRUB), German Centre for Cosmological Lensing (GCCL), 44780 Bochum, Germany\label{aff157}
\and
Department of Physics and Astronomy, Vesilinnantie 5, 20014 University of Turku, Finland\label{aff158}
\and
Serco for European Space Agency (ESA), Camino bajo del Castillo, s/n, Urbanizacion Villafranca del Castillo, Villanueva de la Ca\~nada, 28692 Madrid, Spain\label{aff159}
\and
ARC Centre of Excellence for Dark Matter Particle Physics, Melbourne, Australia\label{aff160}
\and
Centre for Astrophysics \& Supercomputing, Swinburne University of Technology,  Hawthorn, Victoria 3122, Australia\label{aff161}
\and
Department of Physics and Astronomy, University of the Western Cape, Bellville, Cape Town, 7535, South Africa\label{aff162}
\and
DAMTP, Centre for Mathematical Sciences, Wilberforce Road, Cambridge CB3 0WA, UK\label{aff163}
\and
Kavli Institute for Cosmology Cambridge, Madingley Road, Cambridge, CB3 0HA, UK\label{aff164}
\and
Department of Astrophysics, University of Zurich, Winterthurerstrasse 190, 8057 Zurich, Switzerland\label{aff165}
\and
Department of Physics, Centre for Extragalactic Astronomy, Durham University, South Road, Durham, DH1 3LE, UK\label{aff166}
\and
Institute for Theoretical Particle Physics and Cosmology (TTK), RWTH Aachen University, 52056 Aachen, Germany\label{aff167}
\and
IRFU, CEA, Universit\'e Paris-Saclay 91191 Gif-sur-Yvette Cedex, France\label{aff168}
\and
Oskar Klein Centre for Cosmoparticle Physics, Department of Physics, Stockholm University, Stockholm, SE-106 91, Sweden\label{aff169}
\and
Astrophysics Group, Blackett Laboratory, Imperial College London, London SW7 2AZ, UK\label{aff170}
\and
Univ. Grenoble Alpes, CNRS, Grenoble INP, LPSC-IN2P3, 53, Avenue des Martyrs, 38000, Grenoble, France\label{aff171}
\and
INAF-Osservatorio Astrofisico di Arcetri, Largo E. Fermi 5, 50125, Firenze, Italy\label{aff172}
\and
Dipartimento di Fisica, Sapienza Universit\`a di Roma, Piazzale Aldo Moro 2, 00185 Roma, Italy\label{aff173}
\and
Centro de Astrof\'{\i}sica da Universidade do Porto, Rua das Estrelas, 4150-762 Porto, Portugal\label{aff174}
\and
Dipartimento di Fisica, Universit\`a di Roma Tor Vergata, Via della Ricerca Scientifica 1, Roma, Italy\label{aff175}
\and
INFN, Sezione di Roma 2, Via della Ricerca Scientifica 1, Roma, Italy\label{aff176}
\and
HE Space for European Space Agency (ESA), Camino bajo del Castillo, s/n, Urbanizacion Villafranca del Castillo, Villanueva de la Ca\~nada, 28692 Madrid, Spain\label{aff177}
\and
Dipartimento di Fisica - Sezione di Astronomia, Universit\`a di Trieste, Via Tiepolo 11, 34131 Trieste, Italy\label{aff178}
\and
Department of Astrophysical Sciences, Peyton Hall, Princeton University, Princeton, NJ 08544, USA\label{aff179}
\and
Theoretical astrophysics, Department of Physics and Astronomy, Uppsala University, Box 515, 751 20 Uppsala, Sweden\label{aff180}
\and
Mathematical Institute, University of Leiden, Einsteinweg 55, 2333 CA Leiden, The Netherlands\label{aff181}
\and
School of Physics \& Astronomy, University of Southampton, Highfield Campus, Southampton SO17 1BJ, UK\label{aff182}
\and
Institute of Astronomy, University of Cambridge, Madingley Road, Cambridge CB3 0HA, UK\label{aff183}
\and
Space physics and astronomy research unit, University of Oulu, Pentti Kaiteran katu 1, FI-90014 Oulu, Finland\label{aff184}
\and
Center for Computational Astrophysics, Flatiron Institute, 162 5th Avenue, 10010, New York, NY, USA\label{aff185}
\and
Department of Physics and Astronomy, University of British Columbia, Vancouver, BC V6T 1Z1, Canada\label{aff186}}

\abstract{
The matter  around galaxy clusters is distributed over several filaments, reflecting their positions as nodes in the large-scale cosmic web. The number of filaments connected to a cluster, i.e. its connectivity, is expected to affect the physical properties of clusters. 
Using the first \Euclid galaxy catalogue from the \Euclid Quick Release 1 (Q1), we investigated the connectivity of galaxy clusters and how it correlates with their physical and galaxy member properties. 
Around 220 clusters located within the three fields of Q1 (covering $\sim$ 63~$\mathrm{deg}^2$) were analysed in the redshift range $0.2\,<\,z\,<\,0.7$. Due to the photometric redshift uncertainty, we reconstructed the cosmic web skeleton, and measured the cluster connectivity, in  2D projected slices with a thickness of 170 comoving $h^{-1}\,\rm Mpc$ and centred on each cluster redshift, by using two different filament finder algorithms on the most massive galaxies ($M_\star > 10^{10.3}\,\si{\solarmass}$). 
In agreement with previous measurements, we recovered the mass-connectivity relation independently of the filament detection algorithm, showing that the most massive clusters are, on average, connected to a larger number of cosmic filaments, consistent with hierarchical structure formation models. Furthermore, we explored
the possible correlations between connectivities and two cluster properties: the fraction of early-type galaxies and the S\'ersic index of galaxy members. Our result suggests that the clusters populated by early-type galaxies exhibit higher connectivity compared to clusters dominated by late-type galaxies.
These preliminary investigations highlight our ability to quantify the impact of the cosmic web's connectivity on cluster properties with \Euclid. 
}


   \titlerunning{The role of cosmic web connectivity in shaping clusters}
   \authorrunning{Euclid Collaboration: C. Gouin et al.}
   
   \maketitle

\section{Introduction}
The hierarchical model of matter assembly dictates not only the formation and evolution of galaxy clusters, but also their positioning within the large-scale cosmic web, comprising filaments, sheets, and voids. Observations \citep{Joeveer1978,Barrow1985,deLapparent1986,Sousbie2008},
theory \citep{Zeldovich1970,White1979,BKP1996},
and cosmological simulations \citep {Klypin1983,Davis1985,Thomas1992, Springel2005,Angulo2022} have demonstrated that galaxy clusters reside at the nodes of this vast network. They grow through mergers \citep{Bardeen1986,Kauffmann1993} and anisotropic accretion of matter from their connected filaments \citep{BKP1996,Calvo2010,pichon2011,Gouin2017,Vurm2023}. Thus, galaxy clusters serve as unique tracers and laboratories for the study of the structure of the cosmic web, and its environmentally driven influence on galaxy evolution in the densest regions of the Universe.

In the context of cosmic web topology, connectivity---defined as the number of filaments linked to a cosmic node---is one aspect of the broader spectrum of morphological elements \citep[see e.g.][for formal definitions]{Codis2018}. As a key statistical parameter, it probes the geometrical structure of the cosmic web surrounding a cluster.
Currently, increasing numbers of investigations on cluster connectivity are being undertaken to explore the influence of filamentary accretion on the physical and dynamical properties of galaxy clusters, hence yielding better constraints on structure formation. Theoretically, it was shown that mass primarily controls the halo connectivity \citep{Pichon2010,Calvo2010,Codis2018}, such that massive clusters have a larger number of connected filaments than low-mass ones.  Observationally, \cite{Einasto2018B} have shown that groups in superclusters also have higher connectivity than groups of the same richness in voids \citep[see also][]{Einasto2018,Sarron19}. This was  explained from first principle by \cite{Cadiou2020}.
The mass--connectivity relation can be explained by the hierarchical structure formation scenario, in which clusters are the result of merging haloes \citep{Codis2018}. By probing mass assembly history in simulations, \cite{Darragh19} have shown that major merging events increase the connectivity of haloes \citep[see also][for connectivity evolution with redshift]{Lee2021,Galarraga2024}. In agreement with this cluster evolutionary picture, \citet{Gouin21} showed that, regardless of halo mass, highly connected objects grow faster than low-connected ones, linking halo connectivity to its dynamical state captured by its relaxation level. However, this finding is debated by \citet{Santoni2024}, owing to differences in filament-finding methods, cosmic web tracers (galaxies versus gas), and the physics of the underlying cosmological simulations. Regarding halo shapes, cosmological simulations predict that cluster shapes tend to align preferentially with their main connected filaments \citep{Chisari2015,Gouin2017, Okabe2020_SIMU, Kuchner2020, Morinaga2020}, a trend increasingly supported by observational evidence \citep{Einasto2020, Okabe2020_OBS, Gouin2020, Smith2023}.

In addition to its impact on the clusters, the connectivity of haloes is a relevant ingredient for galaxy evolution. At large scales,  \cite{Poudel2017} suggested that cosmic filaments play a role in shaping the properties of groups and their central galaxies. \cite{Darragh19} suggested that high connectivity might be the result of past mergers, which in turn boost the growth of the supermassive black hole and the active galactic nucleus (AGN) feedback \citep{Dubois2013}, and quench the central galaxy.
Interestingly, investigations at smaller scales tend to relate galaxy properties with connectivity. By using both hydrodynamical simulations and SDSS-DR10 observations, \cite{Kraljic2020} showed that more massive, less star-forming, and less rotation-supported galaxies tend to have higher connectivity \citep[see also][]{Tillson2015,Galarraga2023}.

To study the role of connectivity in shaping the properties of clusters and galaxies, it is crucial to reconstruct the skeleton of the cosmic web. However, this requires addressing two key challenges: the definition of filaments and their detection \citep[see e.g.][for a review of filament finders]{Libeskind2017}. Over the past decade, various filament-finding techniques have been developed and applied to both simulations and observations. Among these techniques, \disperse \citep[DIScrete PERsistent Structure Extractor,][]{DISPERSE} and \trex \citep[Tree-based Ridge Extractor,][]{TREX} are widely used on discrete data. In \disperse, filaments are defined as links between maxima and saddles in the density field, using a topological segmentation of the galaxy catalogue that consistently identifies all geometric features of the cosmic web (walls, voids, filaments, peaks). Only the most topologically robust structures are retained. In contrast, \trex defines filaments as a graph-based tree structure modelling the cosmic web skeleton.
These algorithms allow for filament detection in observations. 
As an alternative, the MMF/Nexus formalism  is an explicitly multi-scale method used to analyse  the cosmic web in simulations \citep{Calvo2007,Cautun2014}. 
Dynamically motivated approaches have also been proposed for some time \citep{Arnold1982,Bond1996,Feldbrugge2024} and offer valuable insights into the formation of the cosmic web. However, these methods are less straightforward to implement in observational datasets.

Different studies have tested the capability of \disperse to detect filaments around clusters.  New generations of surveys such as the WEAVE Wide Field Cluster Survey \citep{Kuchner2021, Cornwell2022, Kraljic2022,Cornwell2023}, and \Euclid \citep{EuclidSkyOverview} have motivated a renewed interest. In contrast to detections based on the galaxy distribution, other recent studies have attempted to detect filaments from the gas density field in simulations \citep{Schimd2024,Santoni2024}, and in X-ray observations \citep{DISPERSE,Gallo2024}.

In this context, while cosmic web connectivity is increasingly well understood in Lagrangian space \citep{Codis2018}, large cosmological dark matter simulations \citep{Calvo2010,Codis2012} and hydrodynamical cosmological simulations \citep{Darragh19,Kraljic2020,Lee2021,Gouin21} measuring it in observations remains a challenge. Beyond detecting individual filaments connected to clusters or  superclusters \citep{Einasto2020,Malavasi2020,Aghanim2024,Gallo2024}, large and representative statistical samples of groups and clusters of galaxies with measured connectivities are still rare \citep{Darragh19,Sarron19,Kraljic2020}; to date, this has hindered a full exploration of the impact of cosmic web environments on cluster evolution. 
The situation will change with the ongoing sky survey by \Euclid, which will eventually provide us  with the largest sample of clusters of galaxies containing hundreds of thousands sources \citep{Sartoris2016} and with about ten billion galaxies used to reconstruct the cosmic web skeleton. 

For this work, we used the very first data from the \cite{Q1cite} to explore the connectivity of galaxy clusters as a function of cluster mass and galaxy member properties, and we showcase the capabilities of future analyses of \Euclid data when the survey is completed. To do so, we combined a sample of about 220 already-known clusters, detected in optical and X-ray surveys \citep{ROSAT,SDSS,DES_survey,E_ROSITA}, with the skeleton of the cosmic web surrounding them and derived from the galaxy distribution in \Euclid \citep{Q1-SP028}. In this study, filament finders are applied on a selected sample of galaxies  in 2D projected slices centred on each cluster redshift.
The paper is organised as follows.
In Sect. \ref{sect:data} the selected sample of clusters and the method used to reconstruct the cosmic web skeleton around them are described.
In Sect. \ref{sect:connect} we explore the connectivity of clusters as a function of cluster properties such as cluster mass and galaxy member properties. 
Finally, in Sect. \ref{sect:conclus} we summarise our main results. 
 We consider here a flat $\Lambda$CDM cosmology with cosmological parameters from the {\it Planck} mission \citep{PLANCK_2014}, namely $\Omega_{\Lambda} = 0.693$, $\Omega_{\textrm{m}} = 0.307$, $\Omega_{\textrm{b}} = 0.04825$, $\sigma_{8} = 0.8288$, and $h = 0.6777$.

\section{Cosmic web extraction around clusters \label{sect:data}}

\subsection{Q1 photometric data catalogue}

We present in this section our cluster selection, and the method for extracting the cosmic web skeleton around them in Q1 data, which are divided in three patches: the Euclid Deep Field North of $20 \ \textrm{deg}^2$ (EDF-N), the Euclid Deep Field South of $23 \ \textrm{deg}^2$ (EDF-S), and the Euclid Deep Field Fornax of $10 \ \textrm{deg}^2$ (EDF-F) \citep{Q1-TP001}. In the Q1 dataset, those fields are representative of the Euclid Wide Survey (EWS) in terms of their 5$\sigma$ depths: 26.0, 23.8, 24.0, and 24.0 in the $\IE$, $\YE$, $\JE$, and $\HE$ filters, respectively \citep{Q1-TP005}. Galaxy number counts in the  band for each of the fields are presented in \cite{Q1-TP002}.

\subsection{Cluster selection}

\begin{figure*}
    \centering \includegraphics[width = 0.95\textwidth]{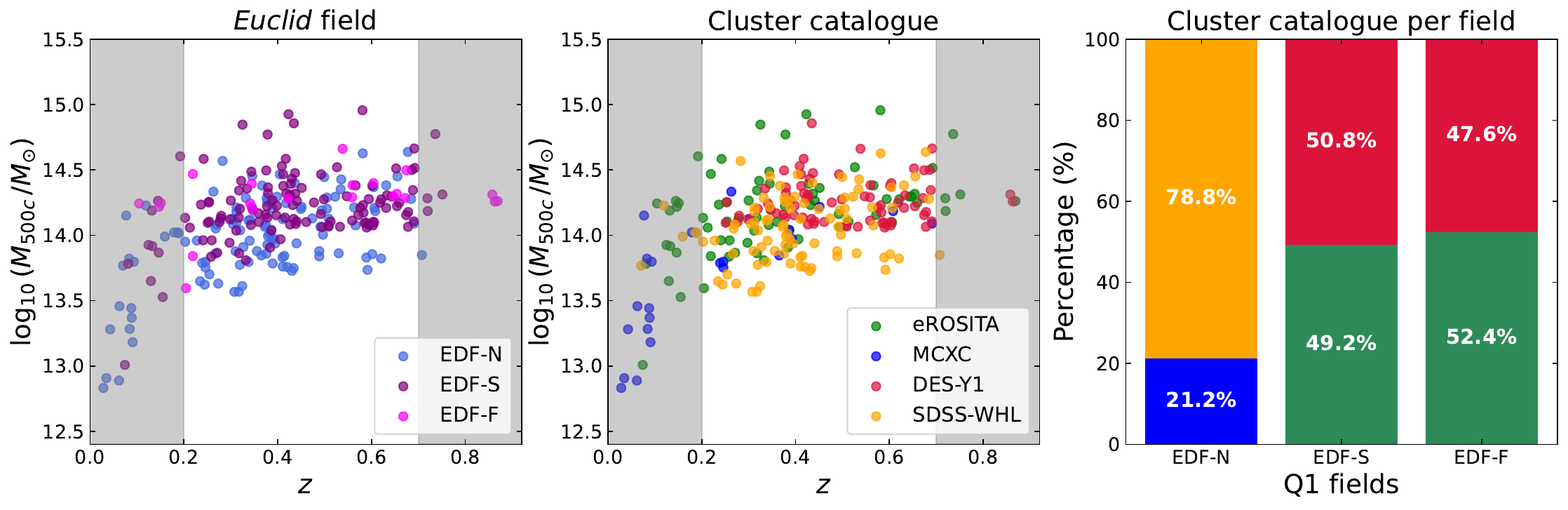}
    \caption{Selected cluster distribution in mass-$z$ space. The clusters are colour-coded according to the \Euclid field they fall in and by their native catalogues. }
    \label{fig:clusters}
\end{figure*}

We focused on the four publicly available catalogues that contain clusters in the \Euclid Q1 fields. These are eROSITA \citep[][]{EROSITA}, MCXC \citep{MCXC}, DES-Y1 \citep{DES}, and WHL-SDSS \citep{WHL2012}. Starting from an initial sample of 322 clusters corresponding to the union of these four catalogues, we finally retained 258 clusters after removing clusters that are excluded by Q1 masks or that appear multiple times in different catalogues. We summarise in Table ~\ref{tab:cluster_samples} the resulting cluster sample and show its properties in Fig.~\ref{fig:clusters}. 
 On the one hand, we estimated the cluster masses from their richness for the DES-Y1 and WHL-SDSS cluster catalogues, following the relation of \cite{McClintock2019} for DES-Y1, and that of \cite{WHL2012} for WHL-SDSS. On the other hand, for the clusters from eROSITA and MCXC, we used the mass values in their catalogues. For each cluster, we estimated the cluster radius $R_{500{\rm c}}$, which is defined as the radius of a sphere that encloses a mass $M_{500{\rm c}}$ with an average density equal to 500 times the critical density of the Universe at the cluster redshift.

In the left and middle panels of Fig.~\ref{fig:clusters}, we show the cluster distribution in mass-$z$ space, colour-coded  according to their location in the three Q1 patches (left panel), and in their initial catalogue (middle panel). As shown in the right panel of Fig.~\ref{fig:clusters}, this cluster sample can be divided into two distinct subsamples, one containing clusters from MCXC and WHL-SDSS located in EDF-N and the second containing eROSITA and DES-Y1 clusters located in EDF-S and EDF-F. We further excluded clusters at redshifts lower than $z = 0.2$, and higher than $z = 0.7$, to avoid bias due to incomplete sampling of the cluster catalogues. Following this redshift selection, our final cluster catalogue consisted of 219 objects.

\begin{table*}[h!]
    \caption{Summary of the selected clusters (see text for details).}
    \centering
    \begin{tabular}{|l||c|c|c|c|c|c|c|c|c|}
    \hline
    Catalogues & eROSITA & MCXC & DES-Y1 & SDSS & Total & Double & Masked & Selected & {$0.2 < z < 0.7$} \\ \hline
    Clusters in Q1 & 76 & 29 & 123 & 94 & 322 & $- 45$ & $- 19$ & 258 & {219} \\ \hline 
    \end{tabular}
    \label{tab:cluster_samples}
\end{table*}

\subsection{Cosmic web detection \label{sect:cosmic_web_detection}}

We discuss here our choice of galaxy selection and the construction of the 2D slices centred on each cluster's redshift performed so that we could extract, from the \Euclid Q1 data, the most robust 2D cosmic web skeletons around clusters. We also discuss the different filament-finder techniques applied on the 2D slices to measure cluster connectivities and their parametrisation. The methodology described here is the same as that presented in \cite{Q1-SP028}. It optimises cosmic web extraction from the Q1 data through Monte Carlo sampling of the galaxy PDF($z$), enabling the construction of tomographic slices.

\subsubsection{Galaxy selection\label{galsel}}

The photometric catalogues from Q1 were produced by the OU-MER pipeline \citep{Q1-TP004}, using images from the VIS and NISP instruments \citep{EuclidSkyVIS,EuclidSkyNISP}, processed by OU-VIS \citep{Q1-TP002} and OU-NIR \citep{Q1-TP003}, along with external ground-based datasets such as the Ultraviolet Near- Infrared Optical Northern Survey (UNIONS).
Morphological parameters, including S\'ersic fits \citep{sersic63}, were measured with \texttt{SourceXtractor++} \citep{Bertin-2020-SourceXtractor-plus-plus}, and additional visual-like galaxy morphologies were obtained using deep-learning techniques \citep{Q1-TP004}. For details on the morphology measurements, we refer to \citet{Q1-SP040} and \citet{Q1-SP047}. The redshifts and stellar masses were derived using the OU-PHZ pipeline \citep{Q1-TP005}, using two distinct methods. The first method, \texttt{Phosphorus}, is based on template-fitting models and provides photometric redshifts along with Bayesian posterior distributions. The second method, \texttt{Nearest-Neighbour Photometric Redshifts (NNPZ)}, is a machine learning-based approach that computes redshifts and stellar masses based on the 30 nearest neighbours from a calibration sample. It provides redshift mode, median, and percentiles.

For the connectivity analysis and to construct redshift slices centred on cluster redshifts, we used both the \texttt{Phosphorus} redshift posterior distribution of galaxies as their redshift probability distribution functions, called PDF($z$), and the \texttt{NNPZ} stellar masses. We start with the galaxy catalogue available in the Euclid Science Archive System and select the galaxies for our subsequent analysis by applying the following steps:  
\begin{itemize}  
    \item Selection of mean photometric redshifts from \texttt{Phosphorus} such that $0 < z < 1$;  
    \item Exclusion of artefacts applying the following quality cuts for retained sources:
        \begin{itemize}  
            \item $\textrm{ \tt phz\_classification} = 2$ (classified as galaxies), 
            \item $\textrm{ \tt phz\_flags} = 0$, $\textrm{ \tt spurious\_prob} < 0.1$, and $M_{\star} < 10^{14}\,\si{\solarmass}$ (free of spurious detections). 
        \end{itemize}  
\end{itemize}

\subsubsection{Construction of 2D slices centred on clusters }

\begin{figure*}[!h]
    \centering
    \includegraphics[width = 0.95\textwidth]{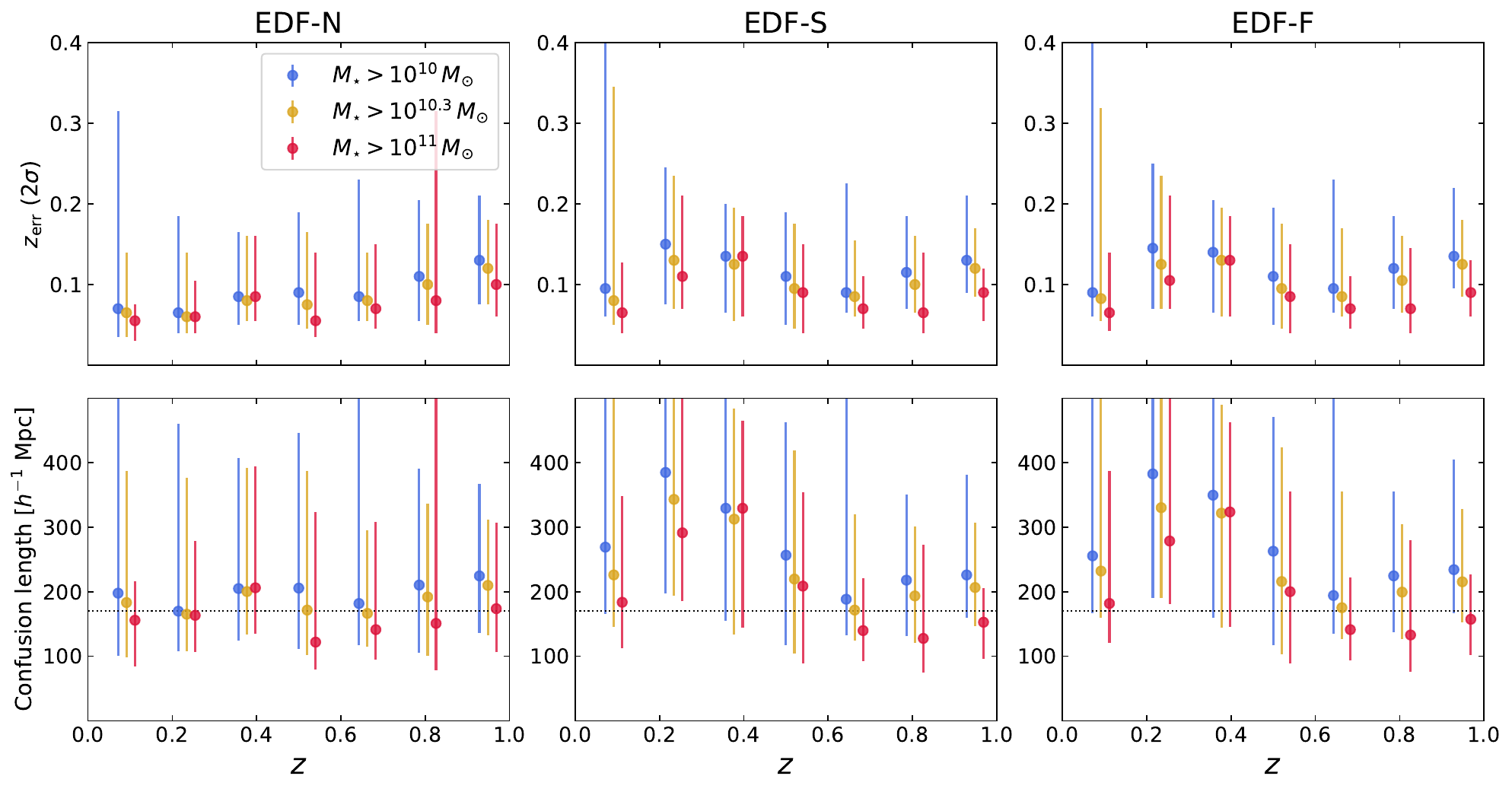}
    \caption{\emph{Top panels:} Median redshift error ($z_{\mathrm{err}} = 2\,\sigma$) as a function of the photometric redshift. We consider here three mass selections of galaxies: $M_{\star} > 10^{11}\,M_{\odot}$ (red), $M_{\star} > 10^{10.3}\,M_{\odot}$ (orange), and $M_{\star} > 10^{10}\,M_{\odot}$ (blue points). 
    \emph{Bottom panels:} Median confusion length, i.e. the associated errors on comoving distance as a function of redshift. The horizontal dotted line represents $170\,h^{-1}~\mathrm{Mpc}$, our choice for the thickness of redshift slices.}
    \label{fig:confusion_length}
\end{figure*}

For each cluster, we construct a 2D slice centred at the cluster redshift in order to detect the 2D cosmic web skeleton based on the galaxy distribution. To achieve this, we first determine the optimal galaxy mass selection and the slice thickness so that the most probable galaxies are included within the cluster slice taking into account the galaxy redshift uncertainties, illustrated in Fig.~\ref{fig:confusion_length} across the three Q1 fields (displayed in separate columns) for various galaxy mass thresholds (colour-coded). The $1\,\sigma$ (high-opacity markers) and $2\,\sigma$ (low-opacity markers) redshift uncertainties are both shown. The top panels of Fig.~\ref{fig:confusion_length} display the redshift uncertainties, while the bottom panels show their corresponding comoving lengths along the line of sight. We conclude that the mean $2\,\sigma$~error is of the order of $170\,\si{\hMpc}$ for galaxies more massive than $10^{10.3}\, \si{\solarmass}$. For each cluster, we decided thus to extract the cosmic web on the 2D galaxy distribution, with galaxies more massive than $10^{10.3}\,\si{\solarmass}$ and projected in slices of thickness $170\,\si{\hMpc}$.

We note that in the northern field (EDF-N), the external data come from the Ultraviolet Near-Infrared Optical Northern Survey (UNIONS), whereas in the southern fields (EDF-F and EDF-S), the data originate from the Dark Energy Survey (DES). These differences naturally explain why the photometric redshift uncertainties are very similar between EDF-F and EDF-S, but are noticeably different in EDF-N. In particular, the absence of $u$-band data in the two southern fields leads to slightly larger photo-z uncertainties at $z < 0.5$. Figure 8 of \cite{Q1-TP001} clearly illustrates the varying depths of these complementary datasets.
Our chosen slice thickness seems small for EDF-S and EDF-F in the redshift range $0.2<z<0.4$. 
 However, since we focus on cluster environments (typically less than $5 R_{500{\rm c}}$), where galaxies are on average more massive, and thus are expected to have more accurate redshifts, this limitation might be mitigated. Enlarging the slice, would significantly degrade the cosmic web reconstruction for EDF-N and other redshift ranges of EDF-F and EDF-S. One possible improvement could be to adapt the thickness to the fields; however,  for the particular case of $0.2<z<0.4$ in EDF-S and EDF-F,  a thickness of $2 \, \sigma$ corresponds to $300 \, \si{\hMpc}$,  and hence is so large that it would enclosed aligned cluster systems.
Following  the literature, we detail the discussion on our choice of 2D slice thickness in the Appendix \ref{app:SLICES}.
 To account for the photometric redshift uncertainties on the population of galaxies belonging to a slice, we performed 100 realisations of each slice by randomly sampling redshifts from the ${\rm PDF}(z)$ of each galaxy. The density of galaxies used to trace the cosmic web varies slightly with redshift, but we did not apply a density cut correction in the present study. However, this will be reassessed with future \Euclid releases \citep[as discussed in][]{Q1-SP028}.
Moreover, because the slice thickness is fixed in comoving units, it corresponds to slightly different multiples of the photometric-redshift scatter in different fields and redshift ranges. This may lead to small field- and redshift-dependent biases that we do not attempt to correct for in this first analysis.

\subsubsection{Filament finder techniques}

To check the robustness  of our results with respect to the filament-finder technique applied in the analysis, we used two different algorithms, \texttt{\disperse} and \texttt{\trex}.
The \texttt{\disperse} algorithm analyses the topology of the density field \citep{DISPERSE2,DISPERSE}. The density field is reconstructed from the discrete galaxy distribution via the Delaunay Tessellation Field Estimator \citep{SchaapWeygaert2000}, where the density is inversely proportional to the area of a triangle in the tessellation. Then, the \disperse algorithm identifies filaments as ridges topologically connecting  pairs of saddle and peak critical points determined through discrete Morse theory. The persistence parameter ($\sigma$) sets the significance threshold for filament detection, effectively distinguishing meaningful structures from Poisson noise. In the context of 2D cosmic web reconstruction with photometric galaxies, \cite{Sarron19} and \cite{Darragh19} have shown that the optimal persistence value for the \disperse algorithm to detect filaments connected to clusters ranges between $\sigma =1.5$ and $2$. 
Following these studies, we set the persistence at $\sigma = 1.5$ and $\sigma = 2$ to capture the large-scale cosmic filaments connected to clusters. These two runs of \disperse, over the 100 realisations of the 219 cluster slices, are discussed further to explore the impact of the persistence on the overall mass-connectivity relation. 

We also utilise the \trex filament finder to detect cosmic web skeletons in 2D slices \citep{TREX,TREX2}. This complements the \disperse findings by providing an alternative method to extract the filamentary structure. The \trex algorithm defines filaments as a set of smooth one-dimensional ridges, leveraging a machine learning extension of the minimal spanning tree where nodes of the graph are represented by Gaussian components of a mixture model. The spatial distribution of these nodes is optimised iteratively using the expectation-maximisation algorithm, which maximises a regularised posterior distribution to best fit the galaxy distribution, while preserving a smooth graph representation and a robustness to uniformly distributed noise in the covered area. The trade-off between accuracy and smoothness is governed by the parameter $\lambda$, which imposes an indirect constraint on the total length of the graph during the optimisation. Graph nodes are initialised using a cut in the extremities of the minimal spanning tree, enabling a proper population of the distribution of galaxies initially. For our analysis, we set $\lambda = 5$ to capture a smooth representation of the large-scale cosmic filaments, while avoiding smaller bridges of matter between galaxies \citep[similarly to][using \trex to compute cluster connectivity in simulations]{Gouin21}. 

To detect the cosmic web skeleton for each cluster, both filament finders are applied to the 100 realisations of their 2D slices. In each realisation, we calculate the cluster connectivity\footnote{With this definition, we measure a multiplicity, and not connectivity according to \cite{Codis2018}.} $\kappa$, defined as the number of filaments crossing a circle of radius $R_k$. In the same range of radial distance as \citet{Darragh19} and \citet{Sarron19}, who  used respectively $R_k = 1.5 R_\mathrm{vir}$ and 1.5 cMpc to measure 2D connectivities in COSMOS and CFHTLS, we measure here the cluster connectivity at radial distances of $2$, $3$, and $4$ $R_{500{\rm c}}$ to investigate possible radial dependence. This yields 100 connectivity values per cluster, from which we compute the mean connectivity and its standard deviation.

\begin{figure*}
    \centering
    \includegraphics[width = 0.33\textwidth]{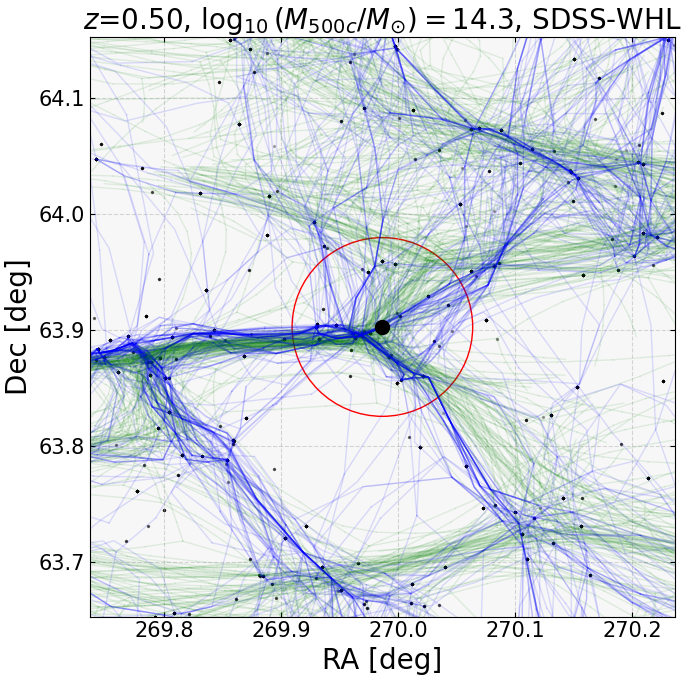}
    \includegraphics[width = 0.33\textwidth]{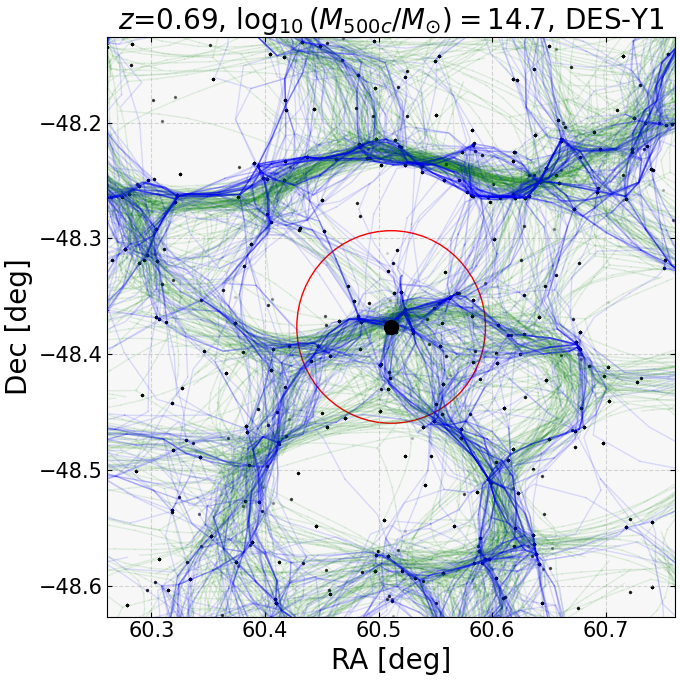} 
    \includegraphics[width = 0.33\textwidth]{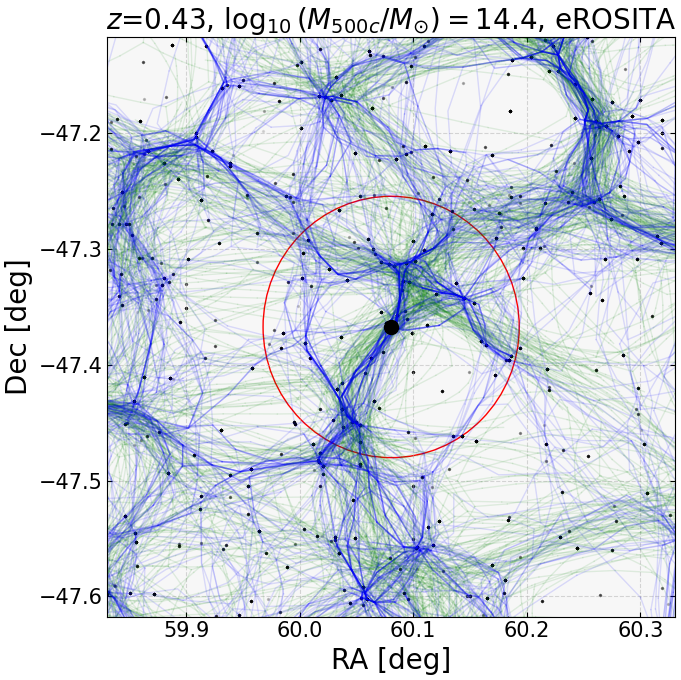}
    \caption{Illustration of the 100 \disperse (blue) and \trex (green) skeletons (from the 100 realisations) found around clusters from the SDSS-WHL (left), DES-Y1 (middle), and eROSITA (right) catalogues. The red circle is centred on each cluster with a radius of $4 R_{500}$. The patches measure $0.5\times0.5 \ \textrm{deg}^2$.}
    \label{fig:TREX_DISP}
\end{figure*}

For illustration, in Fig.~\ref{fig:TREX_DISP} we overlay the 100 2D cosmic web skeletons identified with \disperse (in blue) and \trex (in green) around three clusters extracted from the eROSITA, DES-Y1, and SDSS-WHL catalogues. 
The red circles centred on the clusters show the $4  R_{500}$ radius environments. We observe that both filament-detection techniques agree well and output coherent skeletons. We also note that filament detection is reliable in dense cluster environments, as indicated by the strong opacity resulting from overlapping skeletons across realisations. However, filament identification becomes less reliable in underdense regions. 
As demonstrated in Appendix \ref{app:SLICES}, we recall that the 2D connectivity estimates are only used in a statistical ensemble, and are not intended for the precise characterisation of individual clusters. As illustrated in Fig.~\ref{fig:mock_visu}, we show that 2D filaments, detected in 2D slices with $170\,\si{\hMpc}$ thickness along the line of sight, do not accurately match  the 3D filaments around clusters.

\begin{figure}
    \centering
    \includegraphics[width = 0.48\textwidth]{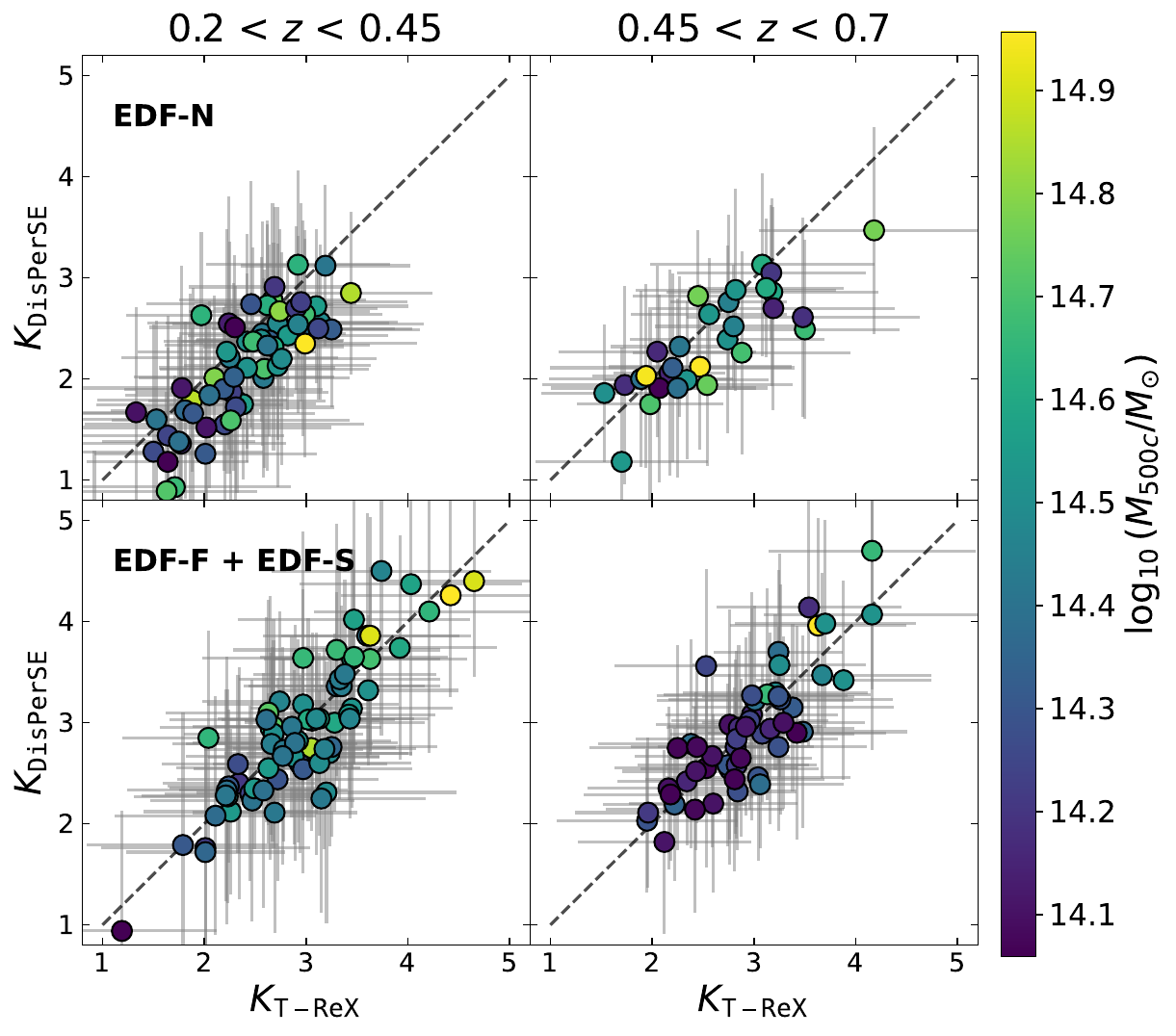}
    \caption{\disperse-\trex connectivity relation  across the Q1 fields. The EDF-N is displayed in the first row and EDF-S+EDF-F in the second row, while the cluster redshifts are presented in the columns. The connectivity points are colour-coded by cluster mass.  }
    \label{fig:K_TREX_DISP}
\end{figure}

In Fig.~\ref{fig:K_TREX_DISP}, we compare cluster connectivity values derived from filaments extracted with \disperse and \trex, showing their one-to-one relation across Q1 fields (rows) and redshift bins (columns). In our redshift range $0.2<z<0.7$, the connectivities from the two algorithms agree very well for all cluster masses and Q1 fields.
We note that in Appendix \ref{app:FILAMENT}, we discuss the tests of different  choices of parametrisation for both \trex and \disperse in order to demonstrate the stability of our results.

\section{Connectivity dependence on cluster properties \label{sect:connect}}

\subsection{Mass-connectivity relation}

\begin{figure}
    \centering
    \includegraphics[width = 0.48\textwidth]{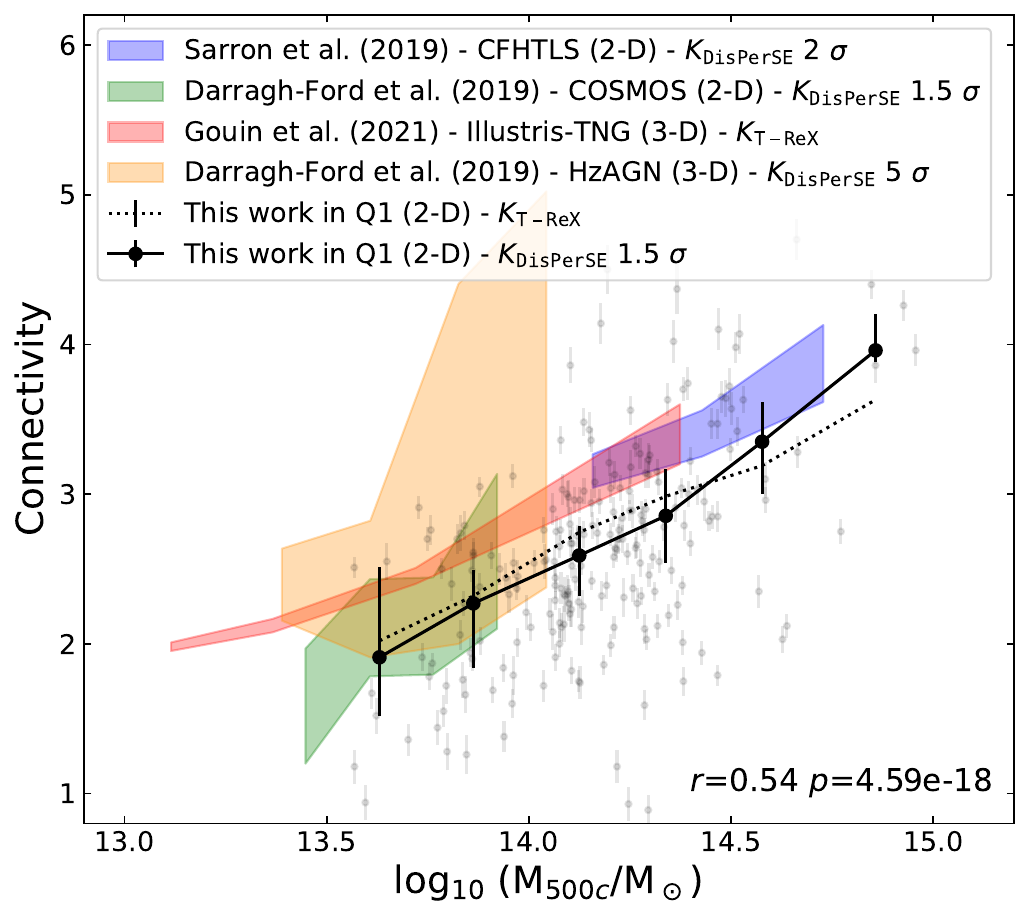}
    \caption{Median mass--connectivity relation measured in Q1 at $0.2 < z < 0.7$ (in black) compared with results obtained, from the Horizon-AGN simulation (in pink;  \citealt{Darragh19}) and the  IllustrisTNG (in red; \citealt{Gouin21}), and with observational results, from CFHTLS  \citep{Sarron19} and COSMOS \citep{Darragh19}. The sigma values in the legend refer to the persistence threshold applied with the \disperse algorithm. 
    The Pearson correlation between cluster connectivity and its mass is given at the bottom of the panel, with  the correlation coefficient ($r$) and the $p$-value.}
    \label{fig:K_mass}
\end{figure}

In Fig.~\ref{fig:K_mass}, we present the median mass-connectivity relation measured in Q1 at $0.2 < z < 0.7$ by using the \disperse algorithm ($1.5 \sigma$ and $R_{k}=4 R_{500{\rm c}}$) alongside 3D connectivity predictions from the IllustrisTNG \citep{Gouin21} and Horizon-AGN \citep{Darragh19} simulations, as well as observational 2D connectivity measurements from COSMOS \citep{Darragh19} and CFHTLS \citep{Sarron19}. Our result provides connectivity measurements over a large mass range,   $M_{500{\rm c}}/M_\odot\in[10^{13.5},10^{15}]$, with a significant correlation coefficient of about 0.5 and in very good agreement with observational measurements from CFHTLS and COSMOS. We note that in this case, we converted the cluster mass given by \citet{Darragh19},\citet{Sarron19}, and \citet{Gouin21}, from $M_{200c}$ to $M_{500{\rm c}}$, by assuming that the ratio $R_{500{\rm c}}/R_{2000c}=0.7$ as estimated by \citet{Ettori2009}.

\begin{figure}
    \centering
    \includegraphics[width = 0.46\textwidth]{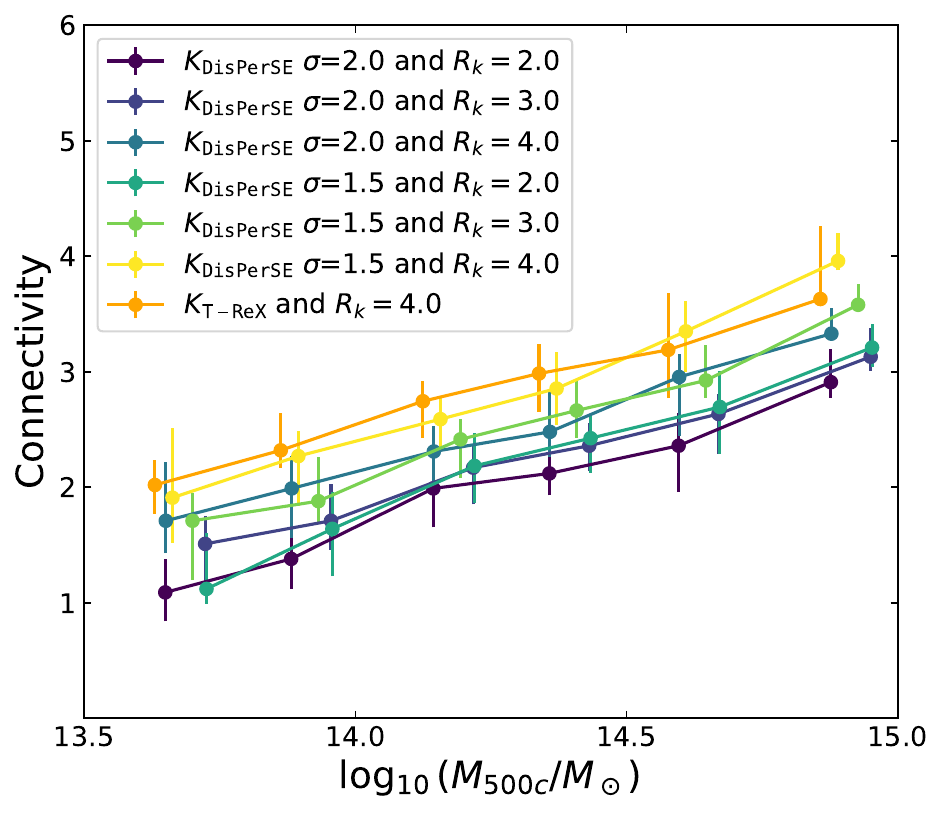}
    \caption{Median mass--connectivity relation measured in Q1 at $0.2 < z < 0.7$, by considering two different \disperse skeletons with a persistence $\sigma=1.5$ and 2, and with three different radial apertures for measuring connectivity $R_k=2,3,$ and 4.}
    \label{fig:K_persitance}
\end{figure}

We recall that the absolute amplitude and the slope of the mass-connectivity relation, $M_{500{\rm c}}$-$\kappa$, are influenced by several factors. First, a key factor is whether the connectivity is measured in 2D projected redshift slices or in full 3D space. As expected, the 2D connectivity measured in Q1 data is slightly lower than the 3D connectivity predicted in simulations. This is explained by  well-known projection effects: some filaments may be aligned along the line of sight or overlap with others.  \citet{Sarron19} and \citet{Darragh19}, who investigated the relation between 2D and 3D connectivity measurements,  show that the 2D photometric skeleton (computed with similar slice thickness) leads to an underestimation of the connectivity compared to 3D connectivities. Related to this, \citet{Laigle2018} showed that the 2D segments of filaments that have no counterpart in 3D are less robust, and thus are removed by assuming a persistence threshold.
Second, in the case of 2D connectivity measurements, the thickness of the redshift slices  varies according to the uncertainties in photometric redshifts. Larger redshift uncertainties lead to thicker slices, which in turn affect the cosmic web reconstruction. 
Third, the choice of tracer used to reconstruct the cosmic web will affect the amplitude of $M_{500{\rm c}}$-$\kappa$ relation. In simulations, cosmic web reconstruction can be performed accurately by using the dark matter or gas density field (providing the detection of small-scale filaments); in observations, filaments are typically detected on the galaxy distributions with a selection cut, such as a stellar mass threshold. Fourth,  the applied filament finder algorithm, its parametrisation, and the radial aperture considered for measuring connectivity might affect the slope of the relation. 
In Fig.~\ref{fig:K_persitance} we present the $M_{500{\rm c}}$-$\kappa$ relation in Q1 at $0.2 < z < 0.7$ using \disperse and \trex by considering different radial apertures ($R_k$) for measuring connectivity and two different \disperse parametrisations ($\sigma=1.5$ and $2$). On the one hand, by increasing the persistence value, we increase the density contrast threshold at which filaments are detected, and thus we reduce the number of detected filaments and cluster connectivity. Conversely, a larger aperture increases the connectivity because it enclose a larger number of filament bifurcations compared to connectivity measurements close to density peaks \citep[as shown by][]{Codis2018}. These listed dependences explain the differences between the $M_{500{\rm c}}$-$\kappa$ relations from the literature in Fig.~\ref{fig:K_mass}. In addition, the scatter of $M_{500{\rm c}}$-$\kappa$ itself is expected to reflect the diversity of cluster mass assembly histories \citep{Cadiou2020,Gouin21}. Therefore, rather than focusing on the absolute amplitude of the $M_{500{\rm c}}$-$\kappa$ relation, we examine in the next section how this relation depends on the physical properties of clusters. This approach provides a more robust means of investigating cluster evolution.

\subsection{Estimation of galaxy members in clusters \label{gal_member}}

To estimate galaxy members within a given cluster, we refined our selection criteria described in Sect.~\ref{galsel}. In addition to the initial cuts applied to exclude spurious sources, we identified galaxy members based on two key conditions.
First, galaxies must reside within the cluster projected sky area, specifically within a projected distance of $2 R_{500{\rm c}}$ from the cluster centre.
Second, galaxies must have a high probability of being near the cluster redshift. Therefore, we computed for each galaxy the probability that it is at the cluster redshift using the general formalism of \citet{George2011} \citep{Castignani_2016,Sarron2021}. In this formalism we model the expected redshift probability distribution of cluster galaxies using a normal distribution, $\mathcal{N}(z | z_\mathrm{c}, \sigma_\mathrm{P})$, centred at the cluster redshift, $z_\mathrm{c}$, and with standard deviation on galaxy photometric redshifts $\sigma_\mathrm{P}$.
We note that this method neglects the uncertainty in the cluster redshift itself, which is appropriate for our data as the typical redshift uncertainty for spectroscopically confirmed clusters is $\sim 0.001 \ (1 + z)$, an order of magnitude smaller than $\sigma_\mathrm{P}$ \citep[see][for a discussion]{Sarron2021}.
Hence, the likelihood of observing a cluster galaxy with $P(z)$ given this model is
\begin{equation}
p(P(z) | {\rm gal} \in C) = \int P(z) \mathcal{N}(z | z_\mathrm{c}, \sigma_\mathrm{P}) \, {\rm d}z \,. 
\end{equation}
Using Bayes’ theorem, we can write the probability that a galaxy belongs to the cluster given its $P(z)$:
\begin{equation}\label{eq:bayes}
    p({\rm gal} \in C | P(z)) \propto {p(P(z) | {\rm gal} \in C) ~ p({\rm gal} \in C)}.
\end{equation}
We consider uninformative priors ${p({\rm gal} \!\in\! C) = 1}$, meaning that in practice we compute the relative probability that the galaxy is at the cluster redshift, assuming the model described for cluster redshift distributions. Following the arguments in \citet{Castignani_2016}, this probability is rescaled such that the maximum achievable probability is one. This is done as in \cite{Sarron2021},
\begin{equation}
    p({\rm gal} \in C | P(z)) = \frac{p(P(z) | {\rm gal} \in C) ~ p({\rm gal} \in C)}{p(P(z) | {\rm gal} \in C, \sigma_P = 0.01)},
\end{equation}
such that a galaxy with photometric redshift distribution $P(z) = \mathcal{N}(z|z_c, \sigma_P = 0.01)$ has a probability of one. We assume that galaxies are identified as cluster members when their probability is higher than 0.5. 
This relative probability threshold is used as a ranking criterion and does not correspond to a fully calibrated Bayesian membership probability; in particular, it should not be interpreted as a literal 50\% probability of cluster membership.
For this initial exploration, we neglected any dependence on magnitudes and radius, such as cluster profiles and the segregation of bright galaxies in the cores.
In future more in-depth analyses, we will use the cluster probability memberships computed by the RICH-CL processing function from the \Euclid LE3 official galaxy cluster detection and characterisation pipeline, which improves on these limitations \citep{Castignani_2016}.

\subsection{Galaxy morphology estimation}

\begin{figure*}[h!]
    \centering
    \includegraphics[width = 0.9\textwidth]{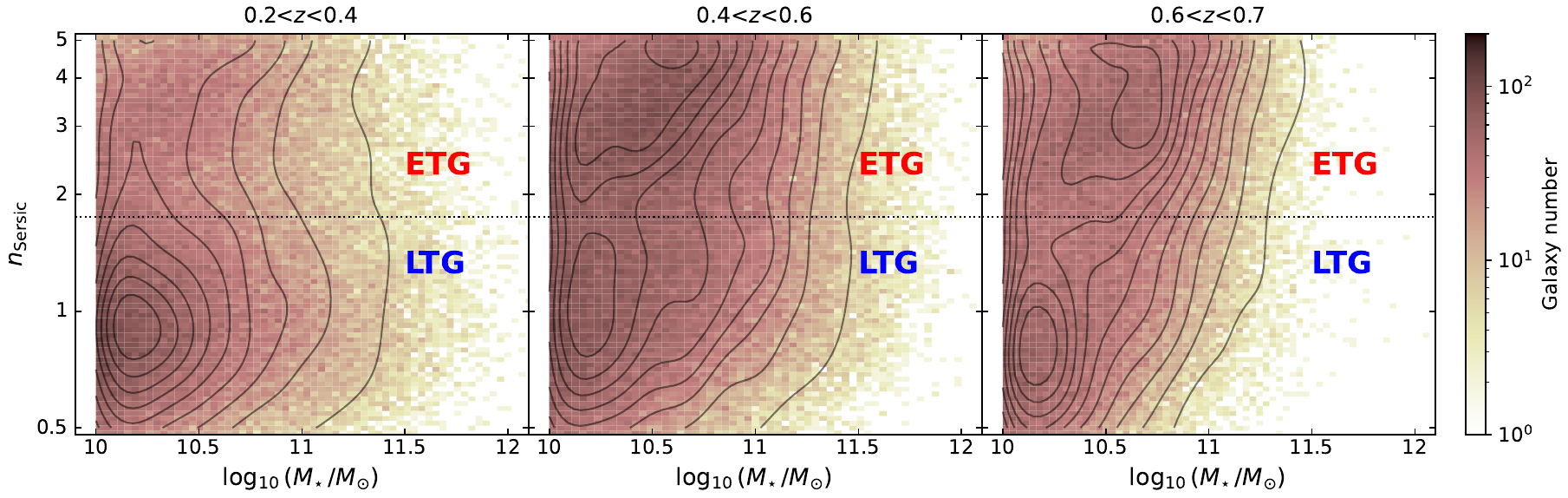}
    \caption{S\'ersic index--stellar mass diagram for galaxies in three different redshift bins: $0.2 < z < 0.4$ (left panel),
$0.4 < z < 0.6$ (middle panel), and $0.6 < z < 0.7$ (right panel). We divided the galaxies into two types: early-type galaxies with $n_{\rm Sersic}>1.75$ and  late-type galaxies with $n_{\rm Sersic}<1.75$.}
    \label{fig:Sersic}
\end{figure*}

For each cluster, the galaxy members are identified following the procedure described in Sect.~\ref{gal_member}. 
We further apply a mass selection such that $M_{\star} > 10^{10.3}\,\si{\solarmass}$, to be similar to the galaxy selection used to trace the cosmic web skeleton (see Sect.\ref{sect:cosmic_web_detection}).
According to \cite{Q1-SP040}, this stellar mass selection should not be affected by mass incompleteness, given that a much more restricted sample with $\IE<23$ is 90\% complete above $M_{\star} > 10^{10}\,\si{\solarmass}$ at $z=0.6$.
Moreover, we explored the galaxy bi-modality  by plotting galaxies according to their location in the $n_{\rm Sersic}$-$M_{\star}$ diagram in Fig.~\ref{fig:Sersic} for the three redshift bins from $0.2$ to $0.7$. 
As shown in this figure, there is a galaxy morphology bi-modality such that early-type galaxies (ETG) are defined by $n_{\rm Sersic}>1.75$ and late-type galaxies are represented with $n_{\rm Sersic}<1.75$.
We note that the $n_{\rm Sersic}$ threshold slightly evolves with redshift, but we ensure that a fixed threshold did not significantly affect our result. 
In general, we found that the fraction of early-type galaxies in clusters is overall consistent with those reported in the literature \citep[e.g.][]{Simard2009}, and is slightly evolving with redshift. Therefore, we later investigated the environmental impact on galaxy member morphologies as depending on both cluster mass and redshift.

\subsection{Relation between connectivity and cluster galaxy morphologies}

\begin{figure*}
    \centering
    \includegraphics[width = 0.99\textwidth]{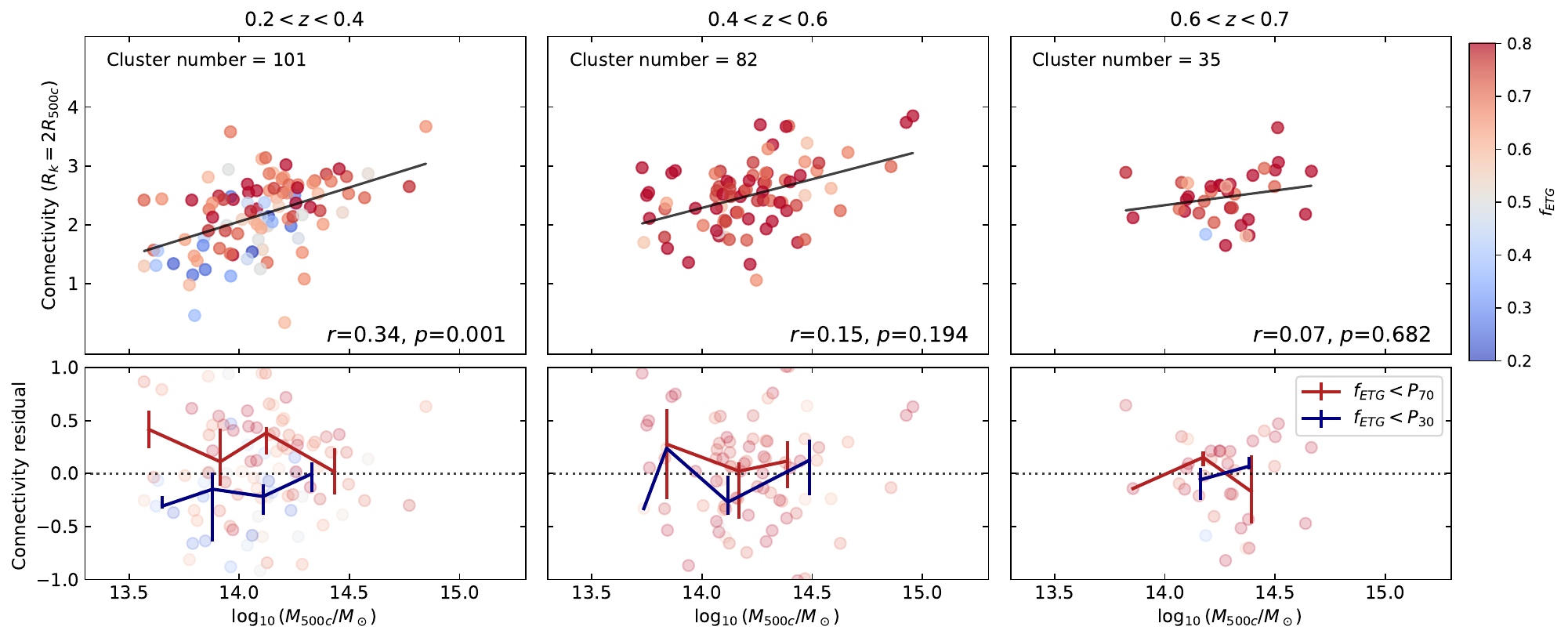}
    \caption{
    Mass--connectivity relation from Q1 data by considering clusters in three different redshift bins: $0.2<z<0.4$ (left panel), $0.4<z<0.6$ (middle panel), and $0.6<z<0.7$ (right panel). The points are colour-coded by $f_{\rm ETG}$ fraction of early-type galaxies inside clusters ($R<2R_{500{\rm c}}$). The black solid lines show the connectivity--mass relation on average, with the connectivities measured at $R=2R_{500{\rm c}}$.
    In the bottom panel we show the connectivity residual, defined as $\kappa- \langle \kappa(M_{500})\rangle$ to remove mass dependence. The red (and blue) solid lines represent the average profile for clusters with $f_{ETG}$ lower than (higher than) the $30{th}$($70{th}$) percentile.
    The partial Pearson correlation between cluster connectivity and ETG fraction, beyond first-order mass dependence, is given at the bottom of the panels, with  the correlation coefficient ($r$) and the $p$-value.}
    \label{fig:K_Sersicz}
\end{figure*}

\begin{figure*}
    \centering
    \includegraphics[width = 0.99\textwidth]{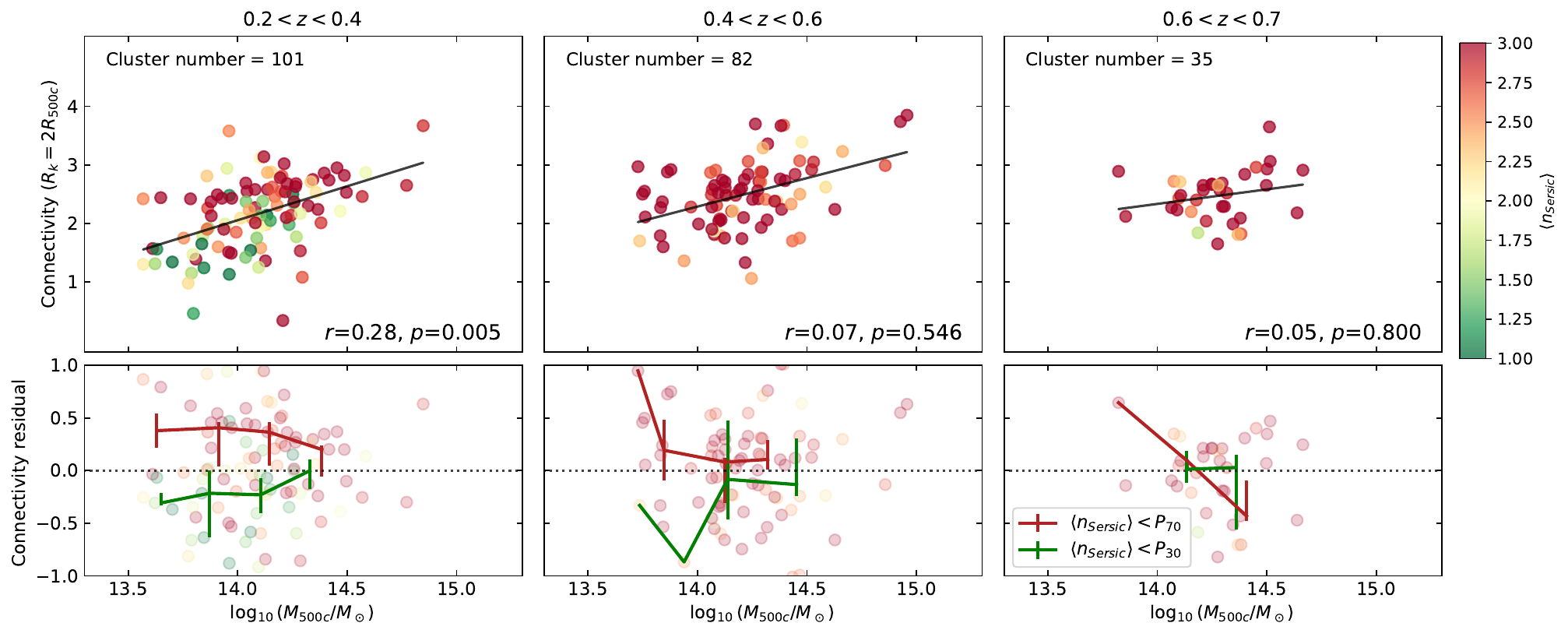}
    \caption{Same as Fig. \ref{fig:K_Sersicz}, but   colour-coded by the median S\'ersic index of galaxies inside clusters $\langle n_{\textrm{Sersic}}\rangle$.}
    \label{fig:K_Sersic2z}
\end{figure*}

By using these methods to characterise cluster members and galaxy morphology, we now explore the relation between cluster connectivity and the morphology of galaxies inside clusters. In Fig.~\ref{fig:K_Sersicz} we present the $M_{500{\rm c}}-\kappa$ relation colour-coded by $f_{\rm ETG}$, the fraction of ETG inside clusters. 
Our cluster sample is divided into three different redshift bins: $0.2 < z < 0.4$ (101 clusters;  left panel), $0.4 < z < 0.6$ (82 clusters;  middle panel), and $0.6 < z < 0.7$ (35 clusters; right panel). We note that, for consistency, we used the same radial aperture of $2 R_{500{\rm c}}$ to identify cluster galaxy members and compute cluster connectivity.
 In addition, to quantify the correlation between the galaxy morphologies inside clusters and the connectivity, beyond mass-driven effects, we used the partial Pearson correlation which measures the degree of association between these two variables, after removing the effect of one (here the cluster mass). 
We found that $f_{\rm ETG}$ tends to correlate with connectivity, only for clusters at $0.2<z<0.4$, with a moderate partial correlation coefficient of 0.34 and a low $p$-value. 
Our result tends to suggest that, beyond the first-order mass dependence, the more a cluster is connected, the more it is populated by early-type galaxies. 
Similarly, we present in Fig.~\ref{fig:K_Sersic2z} the connectivity-mass relation colour-coded by the median S\'ersic index $\langle n_{\rm Sersic}\rangle$ of cluster galaxy members. We can see that, on average, clusters populated by higher S\'ersic index tend to present a higher connectivity. This trend is also only weakly significant for the first redshift bin, with a correlation factor of 0.28 between connectivity and median S\'ersic index.
In Appendix \ref{app:FILAMENT}, we verify that this trend is robust for different \disperse persistence threshold, and we confirm its consistency when using the \trex algorithm with different values of $\lambda$. This appendix therefore demonstrates the stability of our results with respect to variations in filament detection settings.


\subsection{Interpretation and discussion}

Our results appear to be consistent with a scenario in which high connectivity is associated with clusters predominantly populated by early-type galaxies. Supporting this result, \cite{Darragh19} found that galaxy groups with a passive central galaxy tend to have higher connectivity on average than those with a star-forming central galaxy in COSMOS observations \citep[see also][]{Einasto2014}. Analysing hydrodynamical simulations, they suggested that different connectivity levels might trace distinct mass assembly histories, with highly connected groups and clusters having typically undergone their last major merger more recently. Such past merging activity could, in turn, contribute to quenching and morphological transformations of  the central galaxy (and its members). Conversely, it may be easier to preserve late-type morphologies in haloes that have not merged. Additionally, \cite{Kraljic2020} found that less star-forming and less rotation-supported galaxies in SDSS tend to exhibit higher galaxy connectivity, a result further supported by simulations. In highly connected clusters, galaxies are subject to intensified environmental effects, including  tidal and ram-pressure stripping, harassment, and strangulation \citep{Moore1996,Gay2010,Mastropietro2005,Wetzel2013}.
These mechanisms inhibit star formation, leading to galaxy quiescence and a higher proportion of elliptical galaxies.
High-connectivity clusters also tend to be dynamically unrelaxed \citep{Gouin21}, i.e. with higher velocity dispersions, and hence stronger environmental quenching, which leads to the secular  disruption of disk structures 
\citep[see][Appendix D, and reference therein]{Hong2024ApJ}. 
The multiple infalling directions around highly connected clusters may reinforce such mechanisms.
From a theoretical point of view, \cite{Aragon2019} also proposed the Cosmic Web Detachment model, suggesting that as galaxies accrete into filaments, shell-crossing occurs \citep{laigle2015}, cutting off their cold gas supply and ultimately quenching star formation. This model might explain their possible pre-processing, even before they enter clusters \citep[as observed by][]{Conselice_2001,Sarron19, Gouin2020}.
Both pre-processing and processing could explain our findings, weak but statistically significant, that the high fraction of ETGs in clusters is correlated with high connectivity values.

This scenario appears to contrast with several past observational studies that reported enhanced star formation in clusters \citep[see e.g.][]{Porter2007,Fadda2008,Biviano2011,Darvish2014,Lee2019,Ko2024}. However, a number of factors can naturally explain these differences. First, the redshift range probed in some of these works is significantly higher. For example, \citet{Lee2019} analysed eight groups and clusters at $0.6<z<1.3$, i.e. systems in an early assembly phase where cold gas accretion through filaments is still efficient, potentially triggering starbursts during the proto-cluster stage. A similar effect may apply to the $z=1$ sample of \citet{Darvish2014}. Second, the nature of the structures examined varies. \citet{Porter2007} focused on filaments within the Pisces–Cetus supercluster, whereas our study investigates the statistical behaviour of a large population of clusters rather than individual superstructures. Along the same line, \citet{Fadda2008} found that the fraction of starburst galaxies in filaments is more than twice that in the cores of Abell 1770 and Abell 1763. Finally, none of these examples directly quantify connectivity; instead, they rely on proxies for the large-scale environment, such as the friend-of-friend fraction used in \citet{Ko2024}.\\
Taken together, these apparent discrepancies likely reflect differences in redshift, structure type, analysis scale, and galaxy populations considered. Rather than being contradictory, they point towards a more complex evolutionary picture in which (i) proto-clusters may experience an early phase of enhanced star formation, and (ii) supercluster environments or recently merged structures may temporarily boost star formation before quenching becomes dominant. Overall, we argue that the global framework for galaxy and cluster evolution within the large-scale structure is still under development, and that additional simulations and observations will be required to establish a unified scenario.

It should be noted that in this proof of concept investigation we have investigated the dependence of the $M_{500{\rm c}}$-$\kappa$ relation on the morphology of galaxy members. This study serves as a first step towards a more comprehensive analysis;  we will later extend the approach to other galaxy properties such as star formation rate and the fraction of quenched galaxies in clusters. A relation between cluster member morphology and star formation activity has been established in the literature, showing a clear trend in which more massive haloes host a larger fraction of quenched galaxies \citep[e.g.][]{Paccagnella2016, Reeves2021}. However, at this first stage, we focused solely on morphology, as Q1 data provides accurate morphological measurements.

\section{Conclusions \label{sect:conclus}}

This study investigated the role of the cosmic web in shaping galaxy clusters using the first \Euclid Quick Release 1 data.
For this work we used an ensemble of 219 clusters at $0.2<z<0.7$ from the  eROSITA \citep{EROSITA}, MCXC \citep{MCXC}, DES-Y1 \citep{DES}, and SDSS \citep{WHL2012} catalogues. 
By using the photometric redshift posterior distributions of galaxies provided by Q1, we performed a Monte Carlo resampling of the galaxy PDF to build 100 realisations of each 2D slice centred on each cluster. This statistical procedure allowed us to accurately estimate the connectivity for each cluster. By using two different filament-finder algorithms (\trex and \disperse), we ensured the robustness of our connectivity measurements.
{Even if these 2D connectivity measurements are not suitable for the precise characterisation of the cosmic web environments of individual clusters, we demonstrate in Appendix \ref{app:SLICES} that our 2D methodology reasonably recovers statistical information on the connectivity of the large-scale structure in dense environments.

We confirmed the expected mass-connectivity relation predicted by hierarchical structure formation models \citep{Codis2018}. Our result provides 2D connectivity measurements over a wide mass range, and is found to be in very good agreement with past observational measurements \citep{Darragh19,Sarron19}.
Moreover, we explored the relation between the connectivity and morphology of galaxy members. By accurately identifying galaxy members of clusters in Q1 data, we found a moderate correlation, suggesting that the higher the fraction of early-type galaxies, the higher the average connectivity, but only for low-redshift clusters  ($0.2>z>0.4$).
Finally, investigating the median S\'ersic index of galaxy members, we found a weak correlation, indicating that for low-redshift clusters, that the higher the median S\'ersic index of galaxies, the higher the average connectivity. These weak but statistically significant findings, are consistent with a scenario in which high cluster connectivity is associated with clusters predominantly populated by elliptical galaxies. These results are in agreement with the trend found by \cite{Darragh19} on the impact of connectivity on the star formation activity of group central galaxies in COSMOS, and with the results from \cite{Kraljic2020} on relations between galaxy connectivities and their properties.

This work demonstrates the capabilities of Q1 data to investigate the impact of the cosmic web's filaments on cluster evolution. The results pave the way for more comprehensive analyses with future \Euclid data releases, including higher redshift ranges and deeper spectroscopic datasets. 
At the end of the \Euclid mission, the EDF will have been visited 40 times and will provide a  novel spectroscopic sample, including galaxies with an H$\alpha$ flux above $5\times 10^{-17}\,{\rm erg}\,{\rm cm}^{-2}\,{\rm s}^{-1}$ with 60\% completeness. 
In parallel, using the \Euclid spectroscopic sample will allow us to  to reduce the slice thickness (to $25\,\si{\hMpc}$ comoving), and to investigate the galaxy cluster accretion properties in greater detail. 
In a future study, we will  extend the present proof of concept analysis using the catalogue of clusters detected in the Q1 data with their identified galaxy members (Euclid Collaboration: Bhargava et al., in prep.) and later the DR1 \Euclid cluster catalogue.  \Euclid will provide galaxy cluster samples identified using the AMICO \citep{Bellagamba2018,Maturi2019} and PZWav \citep{Werner2023,Thongkham2024} algorithms, extending the overall observed area and reaching higher redshifts, up to $z \sim 2.0$, containing hundreds of thousands of sources \citep{Sartoris2016,Adam-EP3}.

\begin{acknowledgements}
The authors thank an anonymous referee for their useful comments and suggestions. 
 \AckQone
\AckEC  
 We thank St\'ephane Rouberol for the smooth running of the Infinity cluster, where part of the computations was performed.
This research has made use of the SIMBAD and VizieR databases, operated at the Centre de Données astronomiques de Strasbourg (CDS20), Strasbourg, France.
This work has made use of CosmoHub, developed by PIC (maintained by IFAE and CIEMAT) in collaboration with ICE-CSIC. It received funding from the Spanish government (grant EQC2021-007479-P funded by MCIN/AEI/10.13039/501100011033), the EU NextGeneration/PRTR (PRTR-C17.I1), and the Generalitat de Catalunya.
\end{acknowledgements}

\bibliography{aa54655-25} 

\begin{appendix}

\section{Discussion on the slice thickness  \label{app:SLICES}}

The slice thickness results from balancing two competing needs: limiting confusion from stacking filaments of different intrinsic 3D scales, and accommodating photometric redshift uncertainties. This choice is consistent with previous studies that examined projection effects on the 3D filamentary skeleton (Tab.~\ref{tab:appendix}). For instance, \citet{Sarron19} used 300 $h^{-1},\rm Mpc$ slices and \citet{Darragh19} used 120 $h^{-1},\rm Mpc$, both with photometric redshift uncertainties comparable to ours. Using mock catalogues, they showed that although 2D projections lower the absolute connectivity amplitude, the trends with galaxy properties \citep{Sarron19} and with BCG–connectivity correlations \citep{Darragh19} remain robust and physically consistent with their 3D counterparts. \

In addition to previous studies, we test the robustness of our 2D-slice methodology using the \flagship\ simulation \citep{Carretero2017, TALLADA2020}. Specifically, we use galaxies with DR1-like photometry from \flagship\ and apply the same pipeline to a $15 \times 15,\mathrm{deg}^2$ mock of the South field from \cite{Castander2025}. We note that we have performed the same test on the North field mock, but here we show the worst-case scenario, i.e. the South field with higher photo-$z$ uncertainties, to test the cosmic web reconstruction methodology. As shown in Fig.\ref{fig:mock_zerr}, the resulting mock galaxy sample reproduces the photometric redshift uncertainties of the Q1 Euclid Deep Field South data presented in Fig.\ref{fig:confusion_length}. Following Sect.~\ref{sect:data}, we adopt the same galaxy selection ($M_\star > 10^{10.3},\si{\solarmass}$ and $0.1 < z < 0.8$) to construct mock 2D slices of 170 $h^{-1},\rm Mpc$, obtained via Monte Carlo resampling of the galaxy photo-$z$ PDFs.\

To further validate this method, we also construct a 3D map around each cluster using the true redshifts of the mock galaxies. This allows us to obtain the corresponding true 3D skeletons within the same 2D slices. The 3D skeletons are computed 100 times to estimate the intrinsic error on the 3D connectivity: for each iteration, we bootstrap 99\% of the true galaxy distribution and compute the 3D skeleton and connectivity. The mock study contains 1380 clusters, detected with the \texttt{ROCKSTAR} halo finder \citep{ROCKSTAR}, which provides halo masses $M_{\rm halo}$ and radii $R_{\rm halo}$ (dark matter halo properties later used for painting galaxies through abundance-matching techniques). Here, we assume galaxy clusters to be haloes with masses $M_{\rm halo} > 10^{14},M_{\odot},h^{-1}$. We note that, in this mock case, the 3D (or 2D) connectivity is defined as the number of filaments intersecting a sphere (or circle) of radius $R_{\rm halo}$. \

In Fig.~\ref{fig:mock_visu}, we show two examples of mock clusters, with the 2D (blue) and 3D (red) skeletons overlaid on the galaxy distribution around the clusters. True galaxies spatially close to the cluster along the line of sight are marked with red circles. Qualitatively, while the 2D and 3D skeletons appear quite different, they tend to indicate similar directions around clusters.\

In Fig.~\ref{fig:mock_K_mass}, we explore the connectivity–mass relation for both 2D and 3D skeletons. Both capture the same trend with halo mass, suggesting that 2D connectivity can statistically recover the impact of the large-scale environment on cluster properties. We note that the 3D skeleton (persistence of 3) is computed with twice the persistence of the 2D skeleton (persistence of 1.5, the same as in Q1 observations), since the 3D case is essentially noise-free and requires higher persistence to avoid generating filaments at overly small scales \citep[similarly to][for a 2D and 3D \disperse\ comparison]{Darragh19,Sarron19}.\

In Fig.\ref{fig:mock_K2d_K3d}, we examine the correlation between 2D and 3D connectivity and find a Pearson correlation coefficient of 0.28. The substantial scatter highlights the large uncertainties associated with estimating connectivity in 2D. Nevertheless, a positive correlation is still present. To complement this analysis, Fig.\ref{fig:mock_K2d_pdf} shows the probability distribution function of the 2D connectivity for different ranges of 3D connectivity. We observe that the peak of each PDF follows the expected ranking of 3D connectivity, although 2D connectivity is generally lower than its 3D counterpart \citep[as expected from][Fig.~4]{Sarron19}. \

This Appendix analysis demonstrates that our 2D methodology reasonably recovers statistical information on the connectivity of the large-scale structure in dense environments, even though the presence of photometric redshift uncertainties strongly affects the reconstruction of the cosmic web around individual clusters.

\begin{table*}[h]
    \caption{Table of the main studies comparing 2D and 3D connectivity measurements, with predictions applied to mock observations.}
    \begin{tabular}{ l l l l }
    Citations & Observations & Simulations &  2D thickness \\
    \hline
    \hline
    \cite{Laigle2018} & COSMOS2015 & HorizonAGN & 75  $\mathrm{Mpc}.h^{-1}$  \\
    \hline
    \cite{Darragh19} & COSMOS2015 & HorizonAGN & 120 $\mathrm{Mpc}.h^{-1}$ \\
    \hline
    \cite{Sarron19} & CFHTLS & Deep EUCLID lightcone & 300 $\mathrm{Mpc}.h^{-1}$ \\
    \hline
    \cite{Malavasi2025} & - &  GAEA model on Millennium  & 75  $\mathrm{Mpc}.h^{-1}$ \\
    \hline
    EC: Sarron et al. ($in$ $prep$) & - & GAEA model on Millennium  & 25 $\mathrm{Mpc}.h^{-1}$  \\  
    \end{tabular}
    \label{tab:appendix}
\end{table*}

\begin{figure}[h]
    \centering
    \includegraphics[width = 0.42\textwidth]{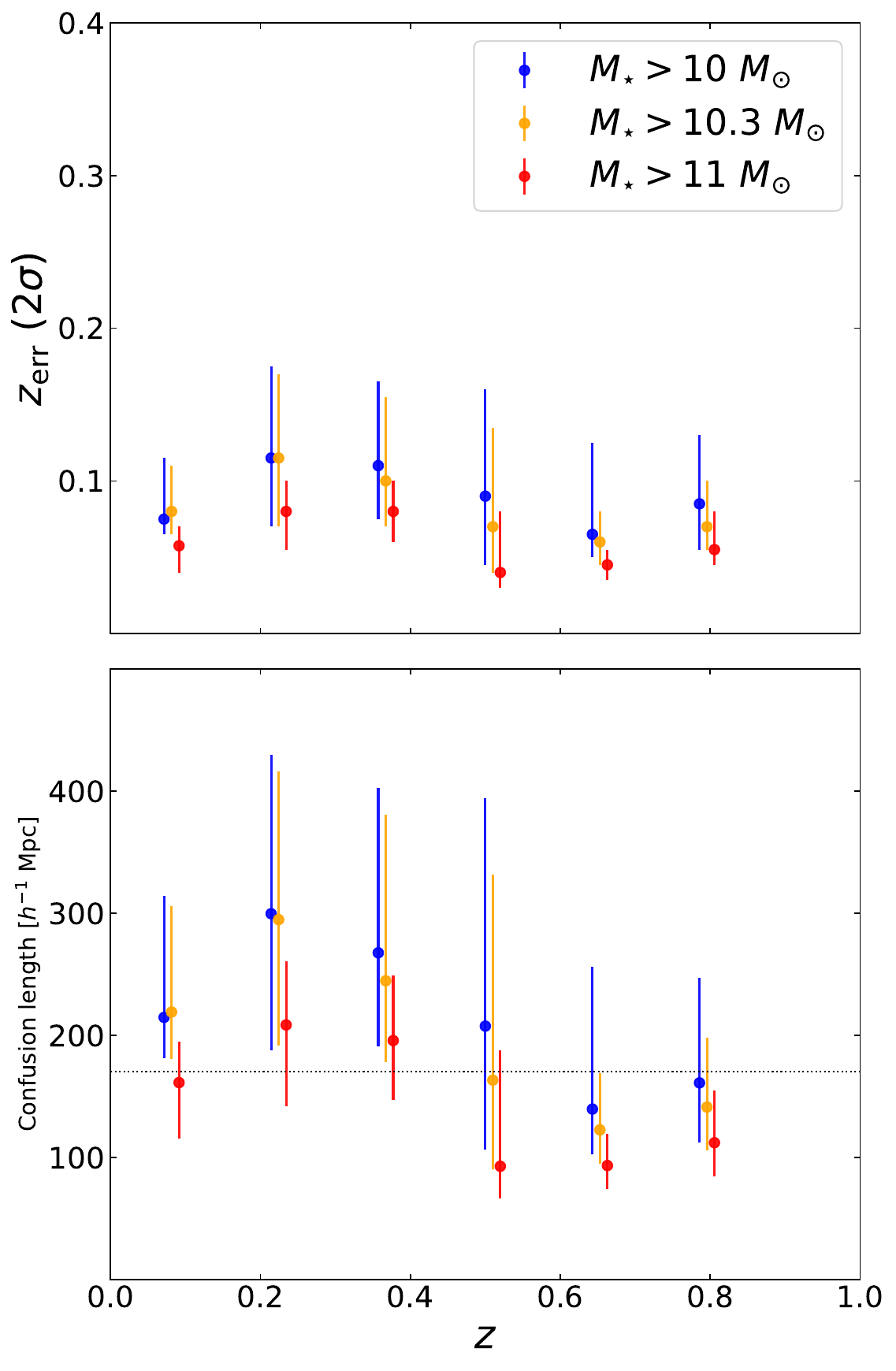}
    \caption{{Same as Fig. \ref{fig:confusion_length}, but considering 
     mock galaxies from the \flagship simulation (using DR1-like photometry).}}
    \label{fig:mock_zerr}
\end{figure}

\begin{figure*}[h]
    \centering
    \includegraphics[width = 0.4\textwidth]{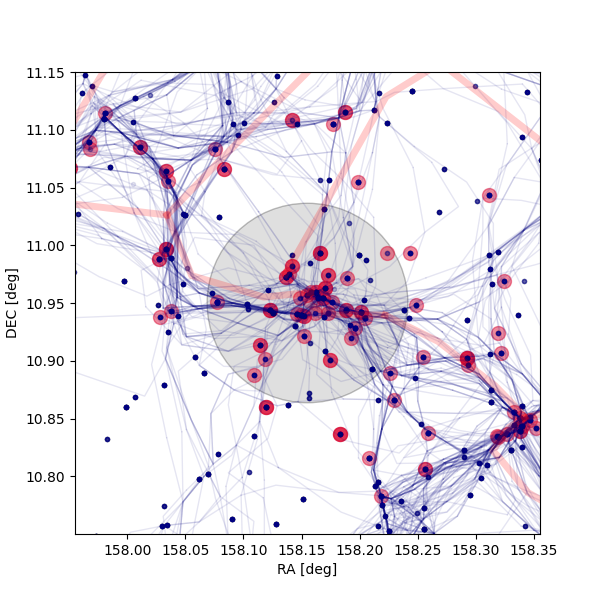}
    \includegraphics[width = 0.4\textwidth]{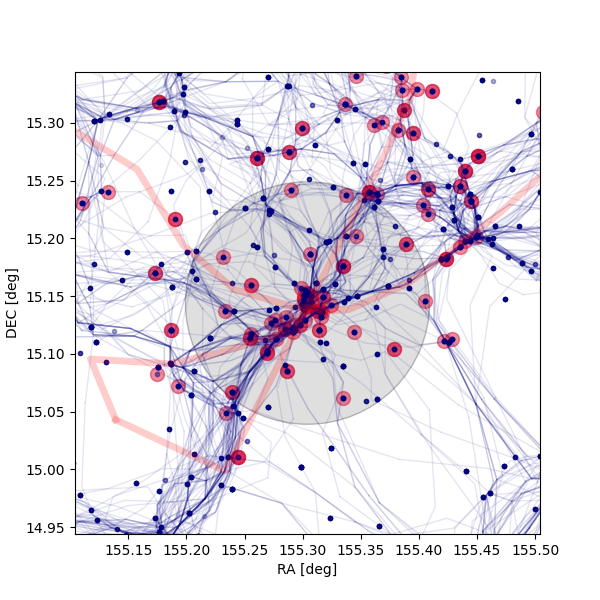}
    \caption{Visualisation of 100 2D skeleton realisations (dark blue lines). Also shown are 3D true skeletons (red lines; projection for 3D filaments between $\pm 3 R_{halo}$). The same galaxies in the 100 realisations of the 2D slices are blue points  and the galaxies that are actually enclosed in $\pm 3 R_{halo}$ along the line of sight are circled in red. The black circles are centred on the cluster halo, with radius equals $R_{halo}$.}
    \label{fig:mock_visu}
\end{figure*}

\begin{figure}[h]
    \centering
    \includegraphics[width = 0.48\textwidth]{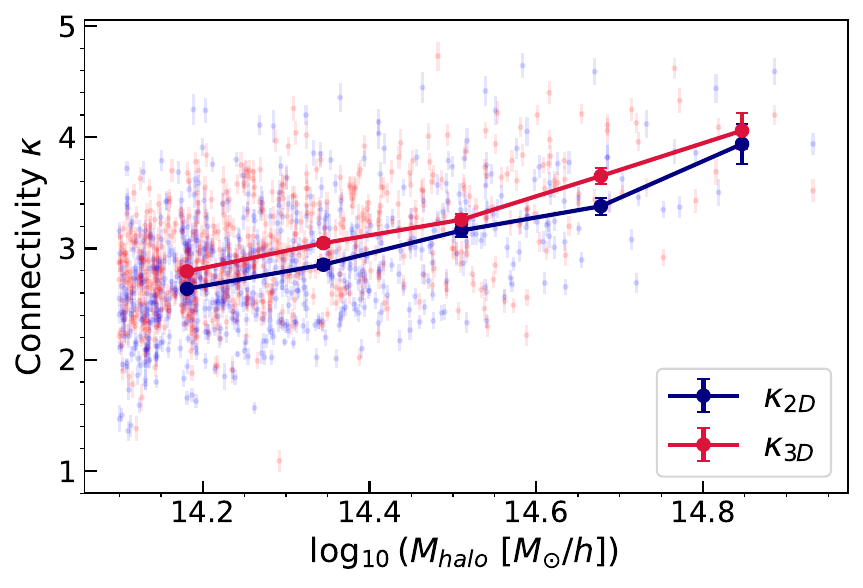}
    \caption{Mass dependence of 2D (red) and 3D (blue) connectivities.}
    \label{fig:mock_K_mass}
\end{figure}

\begin{figure}[h]
    \centering
    \includegraphics[width = 0.46\textwidth]{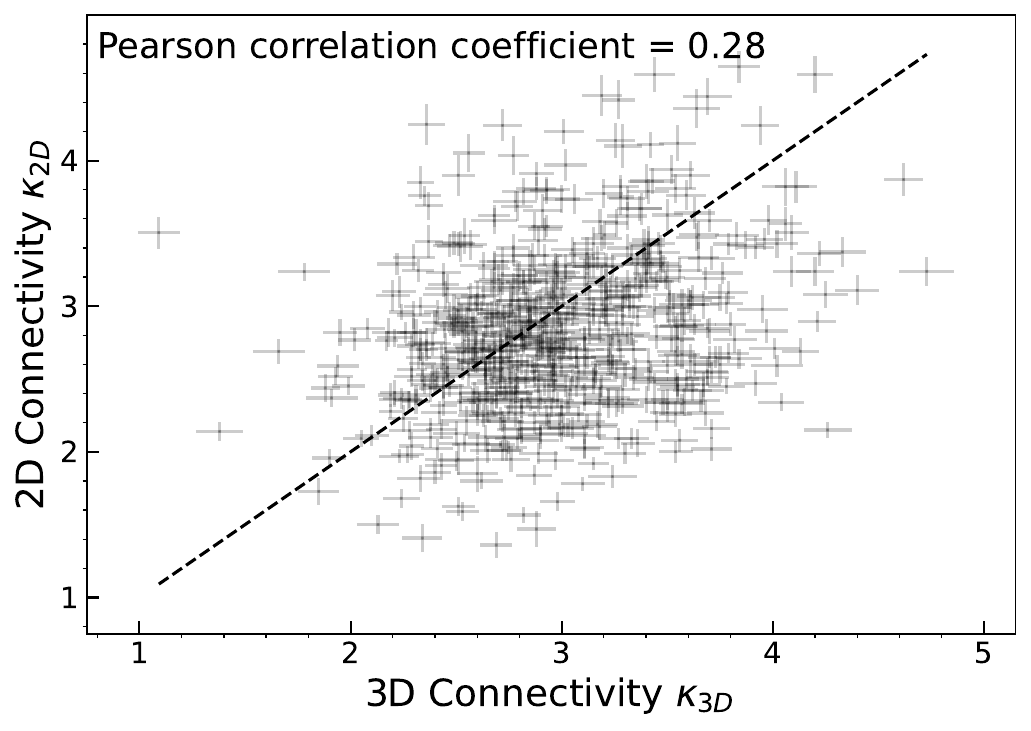}
    \caption{One-to-one relation between 2D and 3D connectivities.}
    \label{fig:mock_K2d_K3d}
\end{figure}

\begin{figure}[!ht]
    \centering
    \includegraphics[width = 0.425\textwidth]{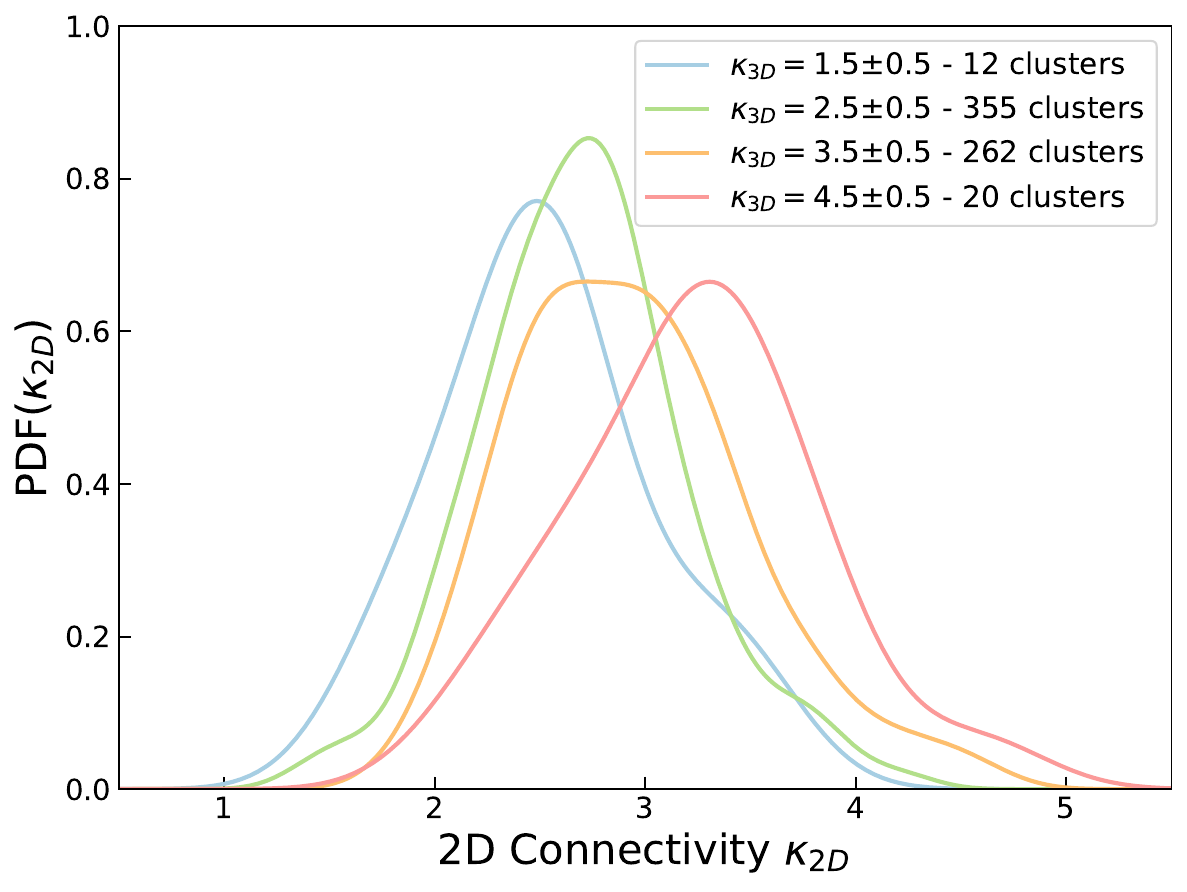}
    \caption{Probability distribution function of 2D connectivity $\kappa_{2D}$ for four ranges of 3D connectivity $\kappa_{3D}$: $\kappa_{3D}=[1-2]$ (blue), $\kappa_{3D}=[2-3]$ (green), $\kappa_{3D}=[3-4]$ (orange), and$\kappa_{3D}=[4-5]$ (red). \label{fig:mock_K2d_pdf}}
\end{figure}

\section{Filament finder parametrisation \label{app:FILAMENT}}

In this Appendix, we discuss the robustness of our results with respect to the parametrisation of both \trex\ and \disperse. Our goal is to verify that the observed trends in cluster connectivity and their correlation with galaxy morphology are not driven by specific choices of parameters in the filament detection algorithms. We focus on the regime where the signal is statistically significant, namely low-redshift clusters ($0.2 < z < 0.4$), in order to assess the impact of cluster connectivity on the morphology of their member galaxies. Cluster member galaxies are identified within $2R_{500{\rm c}}$, and connectivity is measured at $R_k = 2R_{500{\rm c}}$, ensuring a consistent definition of the cluster environment across the sample.\

In Fig.~\ref{fig:q1_disp_test}, we examine the connectivity–mass relation, colour-coded by the ETG fraction, using the \disperse\ connectivity for three different persistence thresholds, $\sigma$: 0.5 (right panel), 1.5 (middle panel), and 2.5 (left panel). We find that skeletons with $\sigma = 0.5$ generally produce noisy filaments, which slightly weakens the observed correlation. In contrast, a persistence threshold of 2.5 yields highly robust filaments but reduces the number of detected structures, thereby narrowing the range of connectivity. Overall, the trend is observed across all persistence values, with $\sigma = 1.5$ providing an optimal balance between noise suppression and filament detection.

A similar test is performed in Fig.~\ref{fig:q1_trex_test} for \trex, where we vary the parameter $\lambda$, which controls the trade-off between accuracy and smoothness in the filament reconstruction. We adopt three values of $\lambda$ (1, 5, and 10), spanning from high sensitivity to more strongly smoothed skeletons. We find that the level of smoothing applied during the filament detection process has a minimal impact on the observed trends between connectivity and ETG fraction.

Overall, these tests demonstrate that our results are stable across different parameter choices for both \trex\ and \disperse. These findings support the robustness of our conclusions, with no strong dependence on the filament detection parameters.

\begin{figure*}
    \centering
    \includegraphics[width = 0.9\textwidth]{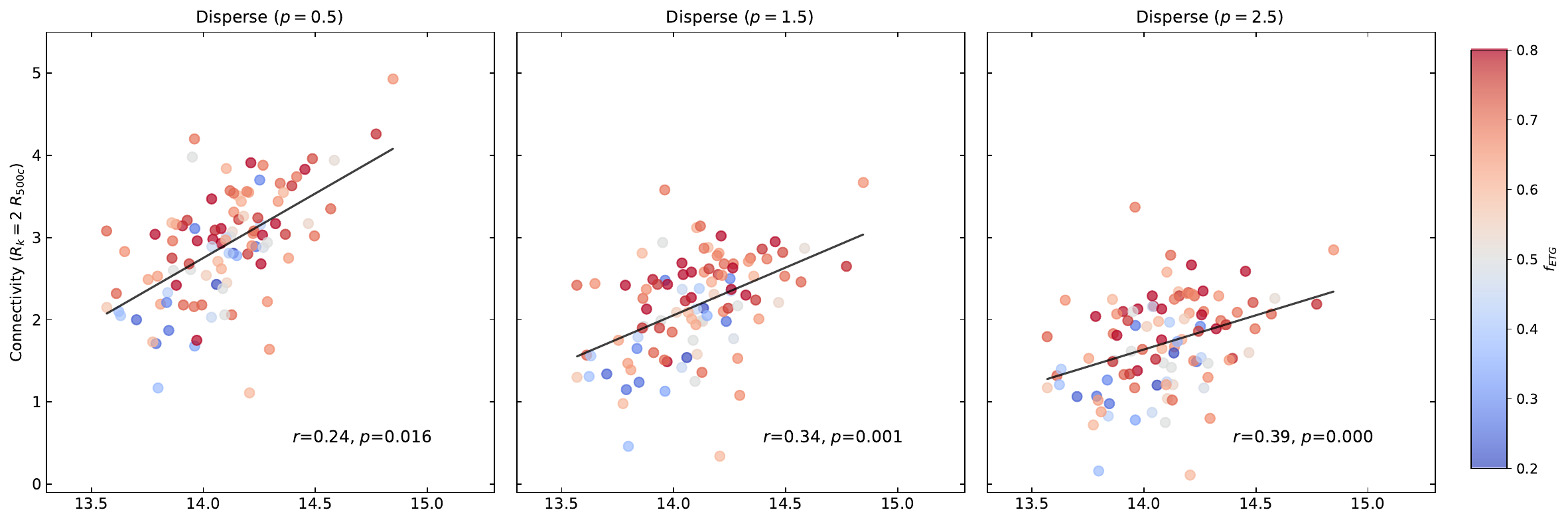}
    \caption{Same as Fig.~\ref{fig:K_Sersicz}, but considering only clusters at $0.2>z>0.4$, and  testing different values of $\sigma$ persistence on our connectivity computed by the \disperse algorithm.}
    \label{fig:q1_disp_test}
\end{figure*}
\begin{figure*}
    \centering
    \includegraphics[width = 0.9\textwidth]{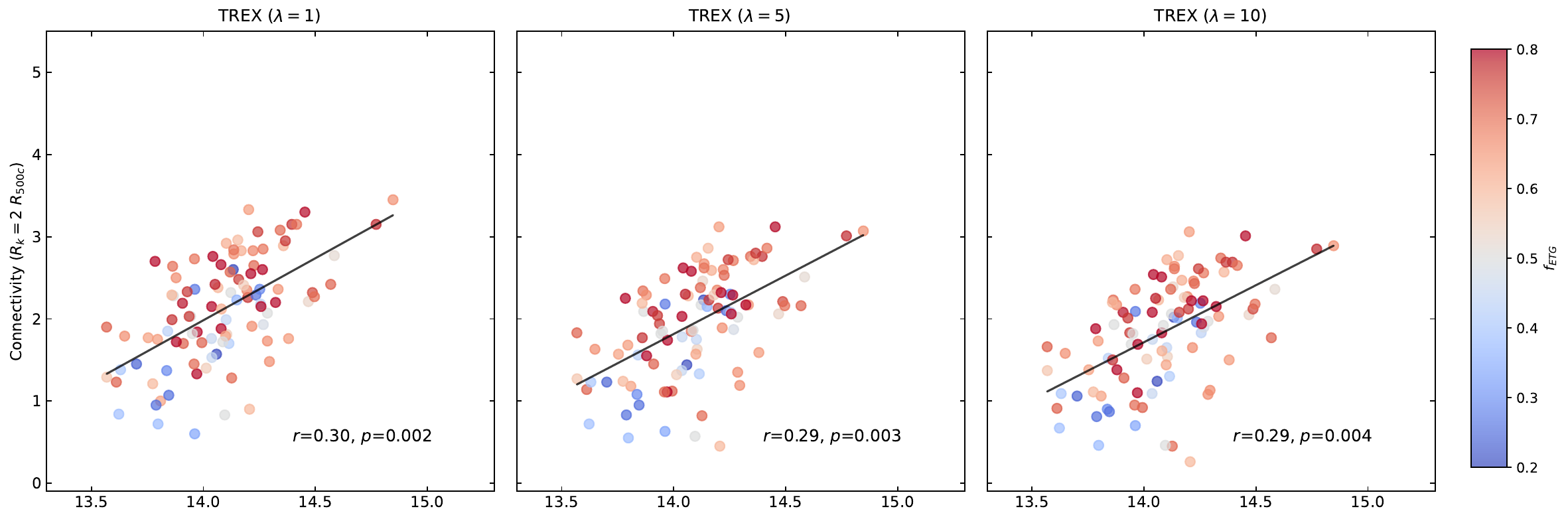}
    \caption{Same as Fig.~\ref{fig:K_Sersicz}, but considering only clusters at $0.2>z>0.4$, and  testing different values of $\lambda$ parameter on our connectivity computed by the \trex algorithm.}
    \label{fig:q1_trex_test}
\end{figure*}

\end{appendix}
 \label{LastPage}

\end{document}